\documentclass[traditabstract,onecolumn]{aa}
\usepackage{graphicx}
\usepackage{natbib}
\usepackage{color}
\usepackage{subfigure}
\usepackage[]{lineno} 
 
\begin{document}
\title{The Fermi Bubbles Revisited}
\author{Rui-zhi Yang\inst{1, 2}
\and Felix Aharonian\inst{1, 3}
\and Roland Crocker\inst{4,1}
}
\institute{Max-Planck-Institut f{\"u}r Kernphysik, P.O. Box 103980, 69029 Heidelberg, Germany
\and Key Laboratory of Dark Matter and Space Astronomy, Purple Mountain Observatory, Chinese Academy of Sciences, Nanjing, 210008, China
\and Dublin Institute for Advanced Studies, 31 Fitzwilliam Place, Dublin 2, Ireland
\and Research School of Astronomy and Astrophysics, Australian National University, Canberra, Australia
}%

\linenumbers
\abstract
{We analyze  60 months of all sky data from the {\it Fermi}-LAT. 
The Fermi Bubble structures  discovered previously are clearly revealed by our analysis.  
With more data and, consequently, better statistics we can now divide each bubble into  constant longitude slices 
to investigate their gross $\gamma$-ray spectral morphology. 
While the detailed spectral behaviour of each slice 
derived within our analysis
is somewhat dependent on the assumed background model, 
we find, robustly,  a relative deficit of the flux at low energies (i.e., hardening) towards the top of the South Bubble.
 In neither Bubble does the spectrum soften with longitude.
The morphology of the Fermi Bubbles is also revealed to be energy dependent: 
at high energies they  are more extended. 
We conclude from the gamma-ray spectrum at high latitudes that a low energy break in the parent cosmic ray population is required in both leptonic and hadronic models. 
We briefly discuss  possible leptonic and hadronic interpretation of this phenomenology. }
\keywords{Gamma rays: diffuse background, ISM:  cosmic rays}
\maketitle

\section{Introduction}
Two huge, bubble-like structures have been reported by \citet{su10}, \citet{dobler10}, and \citet{su12} in {\it Fermi}-LAT gamma ray data to extend $\sim$50$^\circ$ above and below the Galactic center.  
The gamma ray emission from these structures, dubbed the `Fermi Bubbles' (FBs), exhibits a $ E^{-2}$ type spectrum, significantly harder than  the spectrum of the diffuse gamma-ray emission from the Galactic disk.  
Remarkably, structures coincident or similar to the FBs can be seen at other wavelengths, including the  (total intensity) microwave haze found in WMAP  \citep{finkbeiner04} and, most recently, in Planck
data \citep{planckhaze}; the polarized microwave structures reported by \citet{Jones2012}; 
the large-scale, biconical structures found \citep{Bland-Hawthorn2003} in ROSAT X-ray data \citep{rosat}; and the giant polarised radio lobes recently found at 2.3 GHz \citep{carretti13}.  
Several  models  have been proposed to explain both the morphology and spectral properties of the FBs 
\citep{cheng11, mertsch11, crocker11, crocker12, zubovas11, zubovas12, guo2011, guo2012, yang2012,Crocker2013}. 
These models predict different energy dependent morphologies.
For example, in the simplest IC model  (e.g., discussed by Su et al.~2010) with a spatially-independent electron spectrum described by  a power law electron population with a low energy cutoff at $500~\rm$ GeV, one expects   a tendency for softening of the gamma-ray spectra at high latitudes. 
This is due to   the gradual reduction of the  IC  components  produced via  upscattering of the optical and UV photons  - the main target fields contributing to the  production of gamma-rays above $10~\rm$ GeV that also decline towards high latitudes. 
On the other hand, in a simple one-zone hadronic model \citep[e.g.,][]{crocker11} the protons' steady state distribution should produce a similar spectrum of gamma rays at all latitudes. 
Thus, studies of the energy-dependent morphology of the FBs may shed light on the nature of the gamma ray emission mechanism(s).
This is the basic motivation for the current study.

Here we present an analysis based on 60 months' $Fermi$-LAT data. 
We find a similar overall morphology and spectrum for the FBs  to those obtained by \citet{su10}. 
The FBs exhibit a rather homogeneous {\it surface} brightness.
Importantly, this implies  non-homogeneous $volumetric$ emissivities considering projection effects \citep{su10}. 
The FBs, however, are not completely uniform: they exhibit some `hot spots' (referred to by \citet{su10,su12} as the `jet', `donut' and `cocoon'). 
%

In this paper the significantly larger photon statistics and the availability of the recently updated  Fermi science software tools \citep{pass7} allow us  
to investigate the gross spectral morphology of the FBs.
A somewhat similar study has recently been conducted by \citet{Hooper2013}. 
However, they concentrate spectral features in low latitudes while we focus on the  high latitude, especially the top of the bubble. 
%
%
We proceed by dividing each bubble into several slices to investigate 
 possible gamma-ray spectral change with latitude.
Employing different background models, we find, robustly, a spectral hardening, or more specifically,  a deficit of low energy flux  towards the top of the SFB.
We are also able to extend the  spectrum of the FBs to lower energy than previously attempted because of the improvement of the analysis software and the instruments response functions. 
The paper is structured as follows: in section 2 we  present the results of our data analysis, in section 3 we discuss the fitting of leptonic and hadronic models to the gamma ray data, 
and in section 4 we set out our conclusions.

\section{ Data Analysis }
We use the publicly-available data obtained by the {\it Fermi}-LAT  in the gamma-ray energy interval  from $100~$ MeV to $300~$ GeV over the period of  4 Aug 2008 and 17 July 2013 (MET 239557417 - MET 395797613). 
To avoid  contamination from charged particles we use the {\it ULTRACLEAN} data set in the analysis. 
We adopt the instrument response function version P7V6 \citep{pass7}.  
All  events are binned in the all-sky map in HEALPIX format with {\it NSIDE=256}. 
The sources in the $2^{nd}$ Fermi catalog \citep{2fgl} are subtracted from the counts map using the flux in the catalog. 
To take into account the energy dependent point spread function (PSF) of the {\it Fermi}-LAT, we use the  functional form described in the LAT homepage\footnote{http://fermi.gsfc.nasa.gov/ssc/data/analysis/documentation/\\Cicerone/Cicerone\_LAT\_IRFs/IRF\_PSF.html} when masking  sources in the catalog. \\

We use a likelihood method  for determining the FBs' apparent morphology and spectral features for different diffuse emission templates.  
The likelihood function has the form $log(L)=\sum_{k_i}log(\mu_i)-\mu_i-log(k_i!)$, where $k$ is the number of photons in the $i$-th bin in the  counts map and $\mu_i$ is the predicted 
number of photons within a particular linear combination of the templates. 
The sum is over all the spatial bins in the map. 
The  likelihood function is determined in different energy bins to derive  the energy spectrum. 
For fitting  different energy bins, rather than smoothing the map to an universal FWHM (full width at half maximum) as used in \citet{su10}, 
we only smooth the diffuse templates with the Fermi PSF and then fit the counts map as a linear combination of these smoothed diffuse templates.  
The normalization of each diffuse  template is left free in the likelihood fitting. 
We do not assume {\it a priori} the existence of a pair of bubble-like structures. 
We  only use the spatial templates for  $\pi^0$ decay and IC gamma-rays generated by GALPROP \citep{galprop} in addition to the  isotropic template related to the extragalactic gamma ray background and cosmic ray contamination. 
Finally, we take into account the large diffuse structure high above the  plane  to Galactic north which may be connected with the nearby ISM feature Loop 1 \citep{su10}. 
In generating the diffuse templates with GALPROP, we adopt the default Galdef webrun setting, the 2D plane diffusion model that was tuned to fit the ACE data.  

\begin{figure*}
\centering
\includegraphics[width=70mm,angle=0]{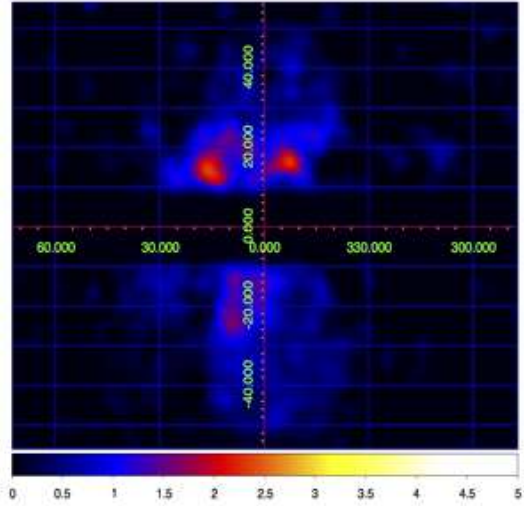}
\caption{The residual map above 2 GeV. Background subtraction is described in the text. Two bubble-like structures can be seen.  To render the picture  clearer we mask the bright Galactic plane $|b|<5^{\circ}$}. 
\label{fig:res}
\end{figure*}

 After subtracting the best-fit linear combination of the diffuse templates, we find  residual maps. 
 By summing over all  energy bins above 1 GeV  we obtain the image in Fig.~\ref{fig:res} in which  two bubble-like structures are clearly seen.  
Next we generate  spatial templates for the FBs from the residual map. 
In the second step of our analysis these Bubble templates are included. 
We employ the likelihood method mentioned above once again and obtain the spectrum of  all the diffuse templates.

\begin{figure*}
\centering
\includegraphics[width=60mm,angle=0]{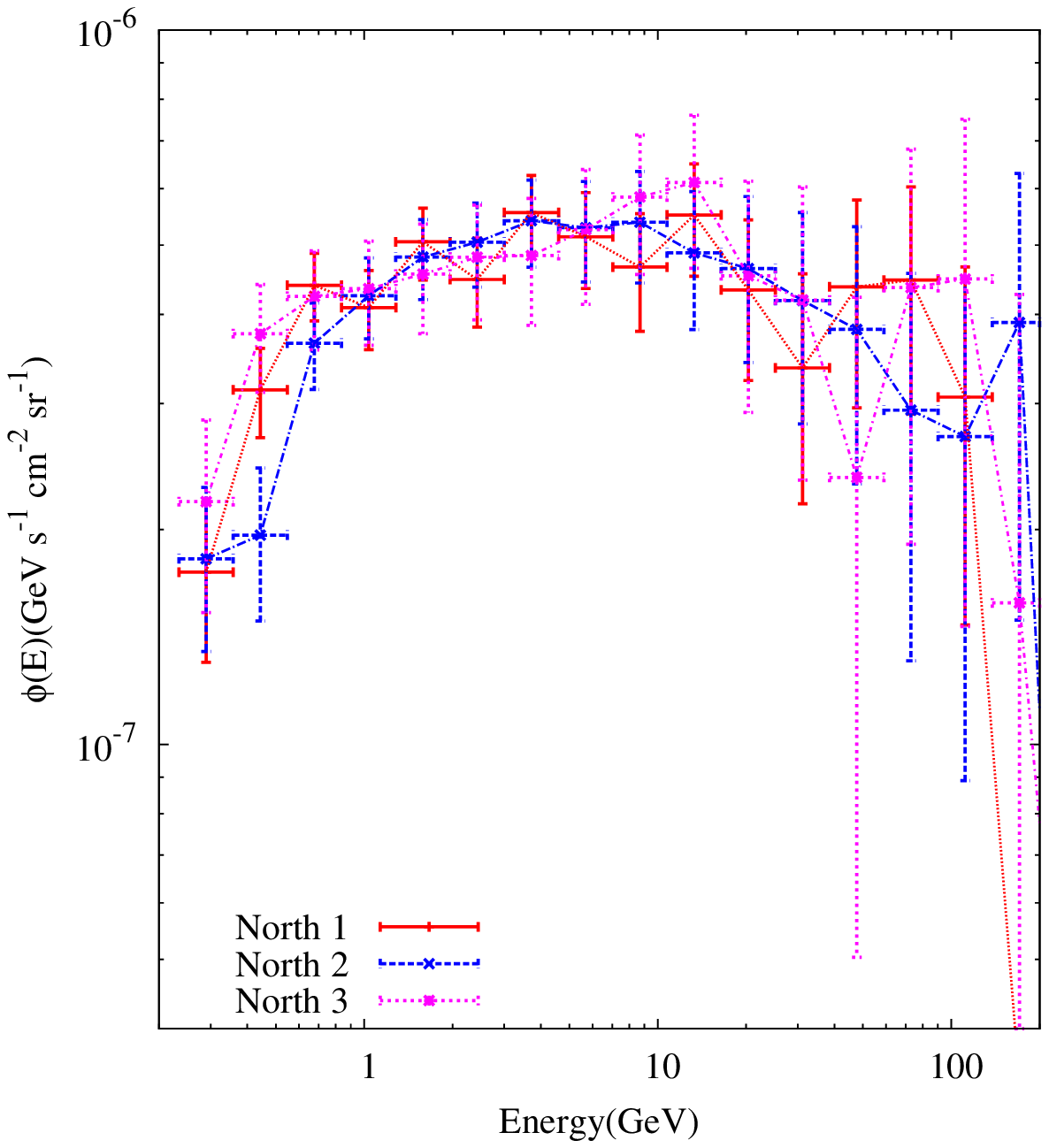}
\includegraphics[width=60mm,angle=0]{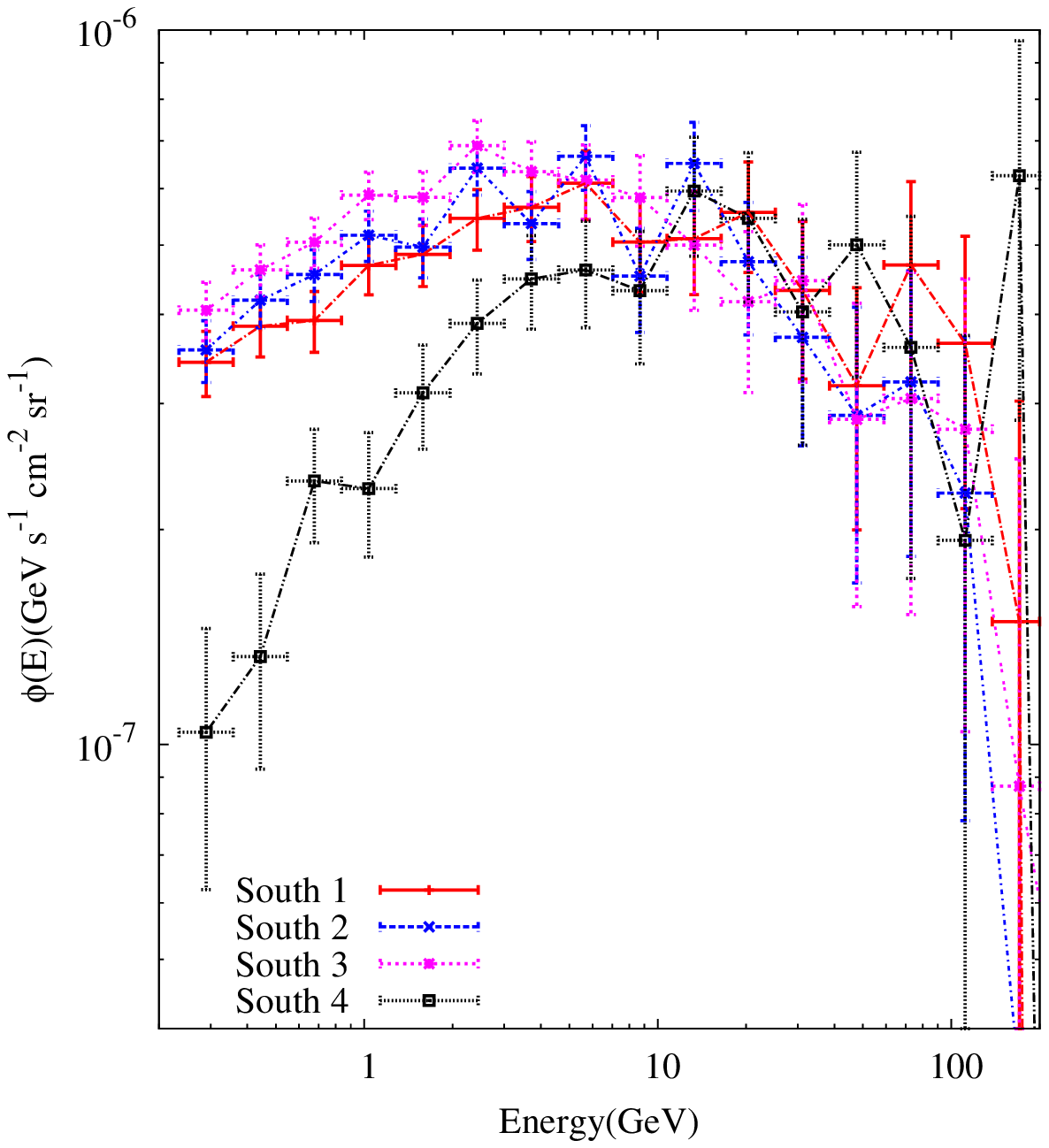}
\caption{SEDs of different slices of the North and South FBs. The numbers 1 to 4 run from low  to high latitudes. Left panel: North Bubble (NFB), Right panel: south Bubble (SFB).  }
\label{fig:sed}
\end{figure*}

\begin{figure*}
\centering
\includegraphics[width=60mm,angle=0]{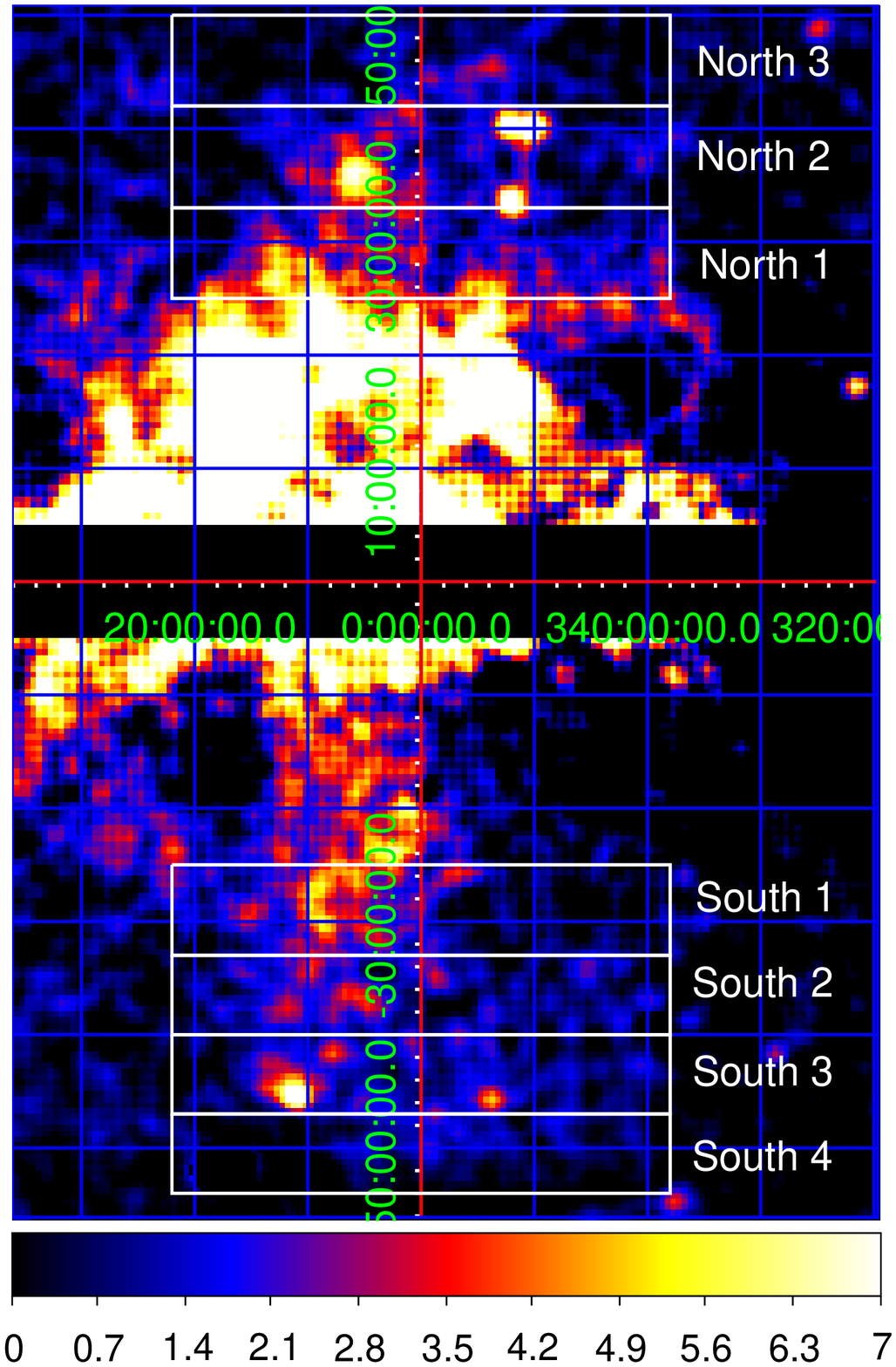}
\includegraphics[width=60mm,angle=0]{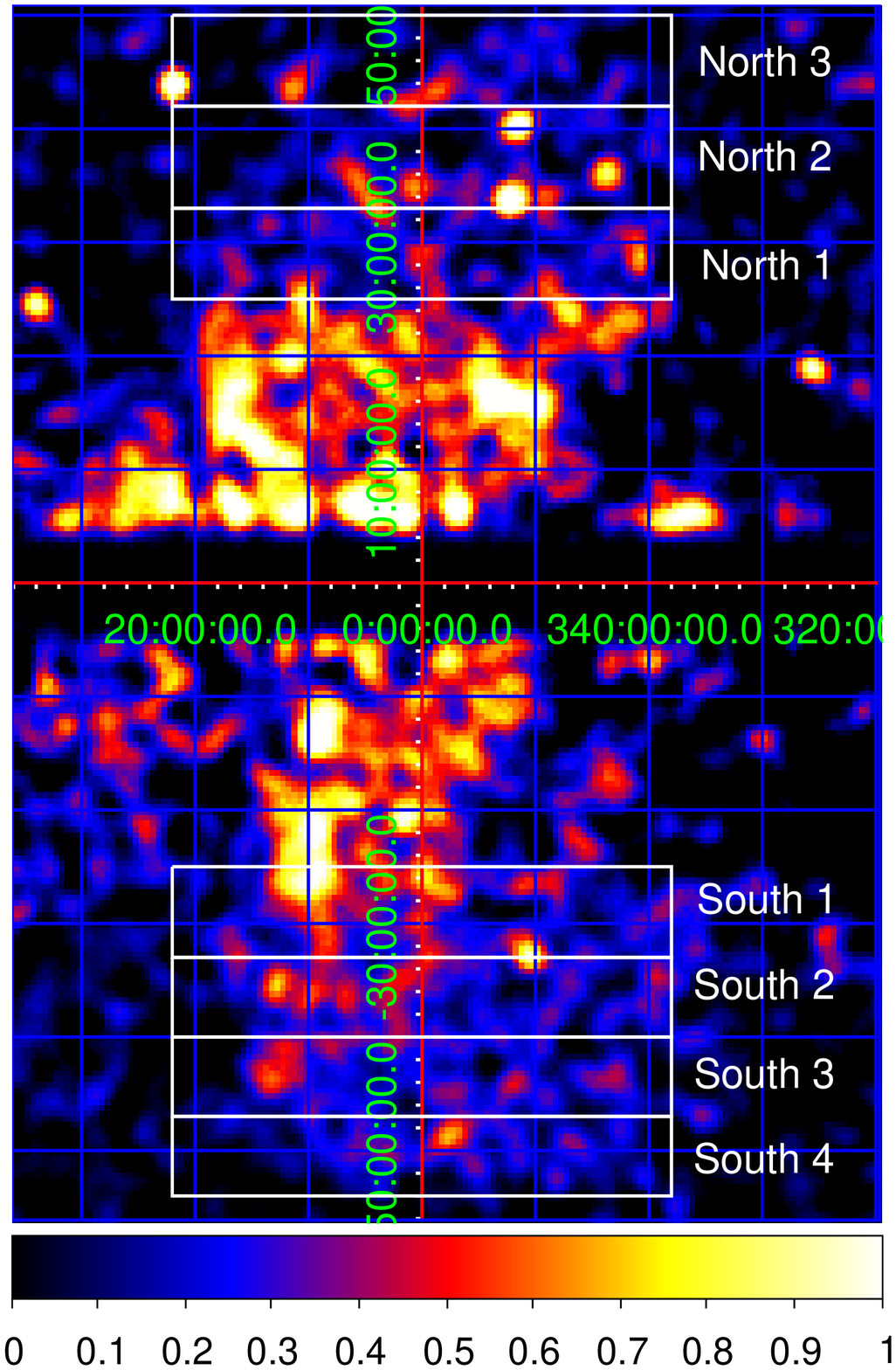}
\caption{Residual maps for different energy bins. The left and right panels correspond to energy intervals $1 - 2$  GeV and $10 - 30$ GeV, respectively. The inner disk of the Galaxy ($|b|<2^{\circ}$) is masked. The boxes laid over the residual maps show the position of each slice.  }
\label{fig:res1}
\end{figure*}

\begin{figure*}
\centering
\includegraphics[width=60mm,angle=0]{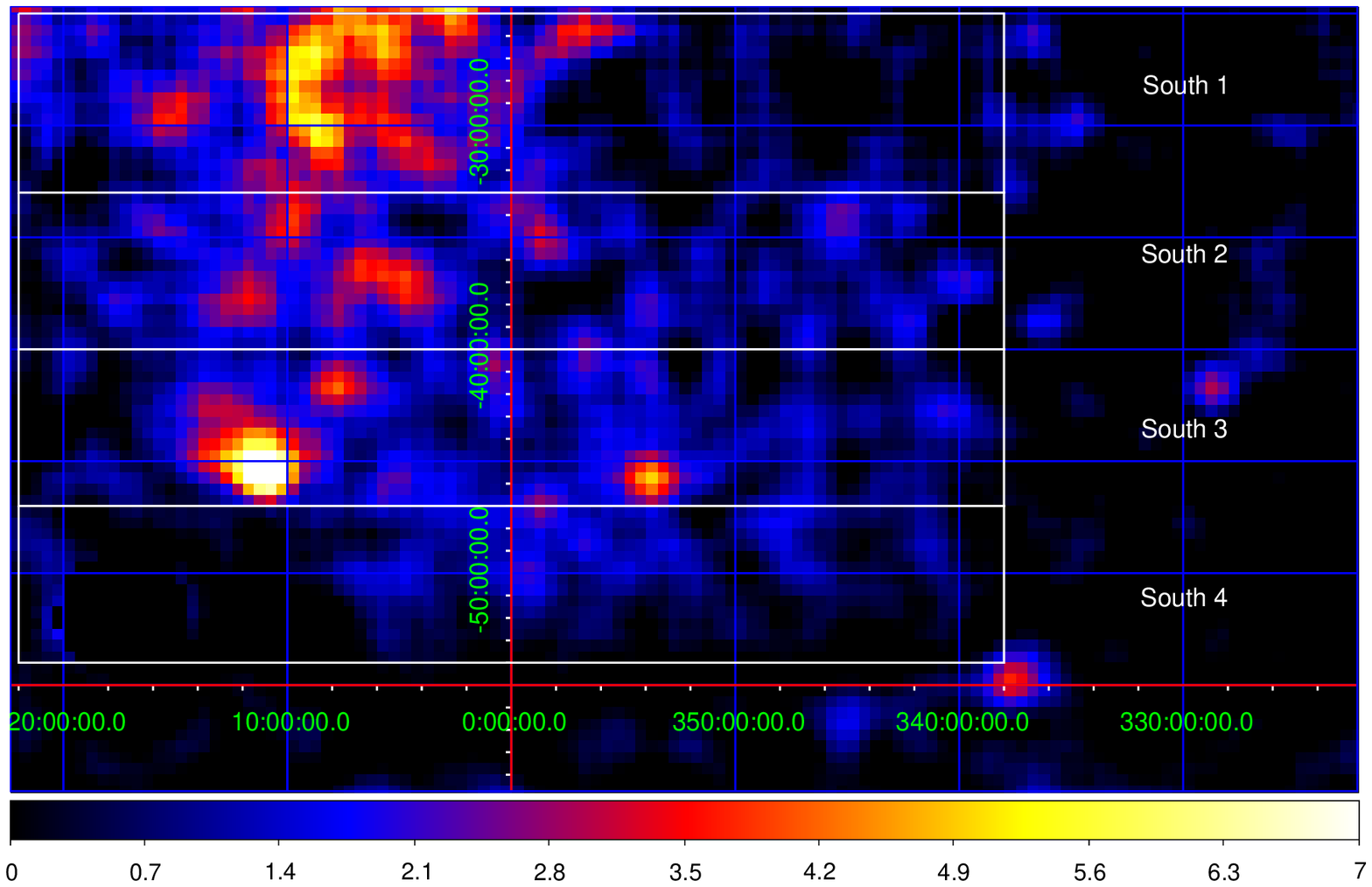}
\includegraphics[width=60mm,angle=0]{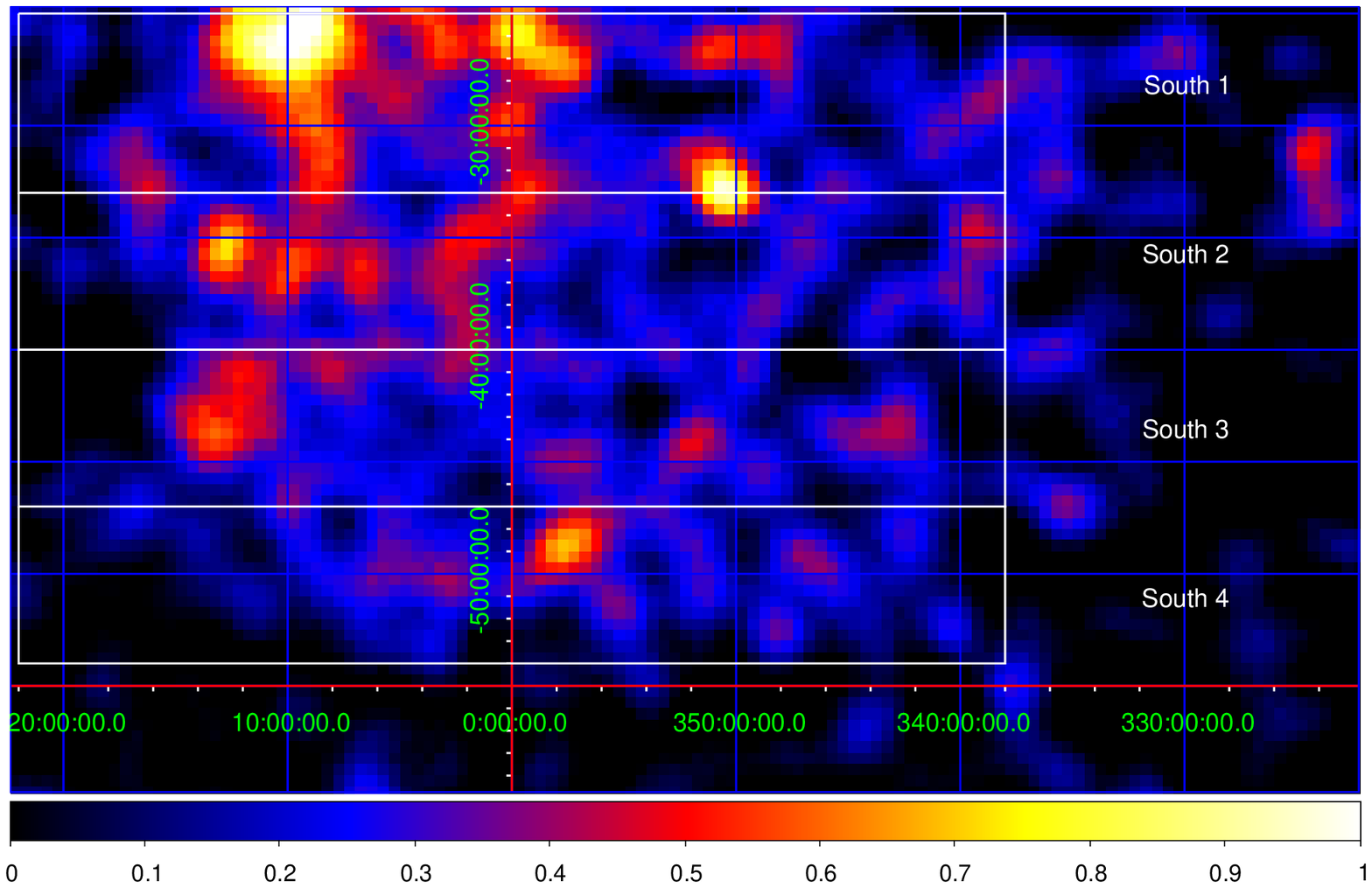}
\caption{Zoom in of Fig.~\ref{fig:res1} for the SFB region. Left panel $1 - 2$  GeV and right panel $10 - 30$ GeV.}
\label{fig:reszoom}
\end{figure*}

To derive the spectrum in different parts of the FBs, we divide the SFB ($-55^\circ < b<-25^\circ$) into four slices and the North Fermi Bubble ($25^\circ <b <50^ \circ $; `NFB') into three. 
To avoid  contamination from the Galactic plane, in the fitting we  mask out the inner $\pm 25^\circ$ region. 
The position of each analysis slice can be found in Fig.~\ref{fig:res1}.   
The SED of each slice is shown in Fig.~\ref{fig:sed} where the numbers 1 to 4 run from  lowest  to highest latitude. 
From the SEDs it is evident that the spectrum of the highest southern slice is in  deficit at low energies relative to the other slices.
It is important to note that this same slice suffers little from geometric projection effects and  is thus a true reflection of the spectrum at the top of the SFB. 
The spectrum at the top of the SFB is, therefore, significantly different from that in the interior.   
 Fig.~\ref{fig:res1} reveals the different morphologies of the FBs in different energy bins; at high energies the SFB is clearly more extended than  at low energies.
This is highlighted  in Fig.~\ref{fig:reszoom}. 
It should also be noted that in the top slice of the SFB (South 4) there are no known point sources from the $2^{nd}$ Fermi catalog,  
making our conclusion about the spectral variation quite robust.

Regarding the energy-dependent morphology of the FBs, we also flag the following point to be dealt with in further work: 
as evident from figs.~\ref{fig:res1} and \ref{fig:reszoom}, the SFB is also relatively more extended to Galactic {\it west} at high energies than low energies implying a spectral hardening going from east to west.
As far as we are aware, no model for the FBs currently accounts for this effect.
It is interesting that, because of this extension to the west, at high energies, the FBs come to more closely resemble the
polarised radio lobes detected at 2.3 GHz \citep{carretti13}\footnote{RMC thanks Ettore Carretti for raising this point in conversation.}.

Accurate determination of the morphology and spectral features of the Bubbles strongly depends on reliable  modeling of the diffuse background. 
Unfortunately, because of uncertainties in the distributions of cosmic rays and  interstellar gas, our knowledge about the diffuse gamma-ray background is imperfect. 
In such circumstances, one can only introduce different {\it templates} that instantiate  different estimates of the background and try to derive the true Bubble flux with them using a likelihood method. 
Unfortunately, different choices of background templates  alter the final result significantly and this may lead to significant systematic errors. 
To study such possible errors, we investigate the 128 GALPROP models listed by \cite{fermidiffuse}. 
The aim of \citet{fermidiffuse} was to study the origin and propagation of cosmic-rays and the distribution of the interstellar medium  by simultaneously fitting  diffuse gamma ray emission and  cosmic ray data. 
Although all their  models underestimate the GeV emission at low latitudes \citep{fermidiffuse}, this has little influence on our results given our masking of the inner $\pm 25^\circ$ when deriving spectra.
The 128 models are different in cosmic ray (CR) source distribution, CR halo size, HI spin temperature and $E(B-V)$ magnitude cut. 
Following perusal of the the online material for  \citet{fermidiffuse}, we found that 64 models with $z=8 ~\rm kpc$ and $z=10 ~\rm kpc$ do not fit the $^9Be/^{10}Be$ data well (the derived curves fall outside of the error bars of nearly all the data points),  where $z$ is the height of the CR halo.  
Therefore we abandon these, and investigate the other 64 models as diffuse emission templates. 

\begin{figure*}
\centering
\includegraphics[width=60mm,angle=0]{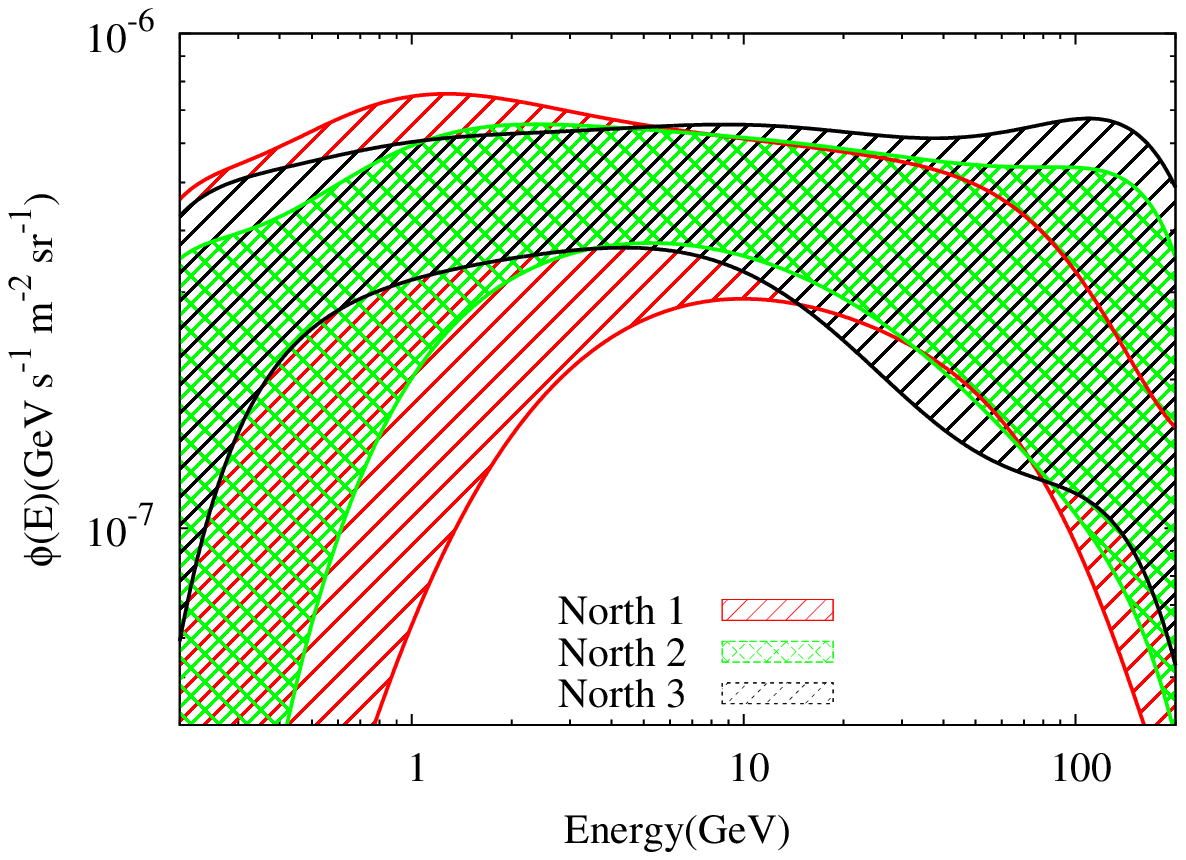}
\includegraphics[width=60mm,angle=0]{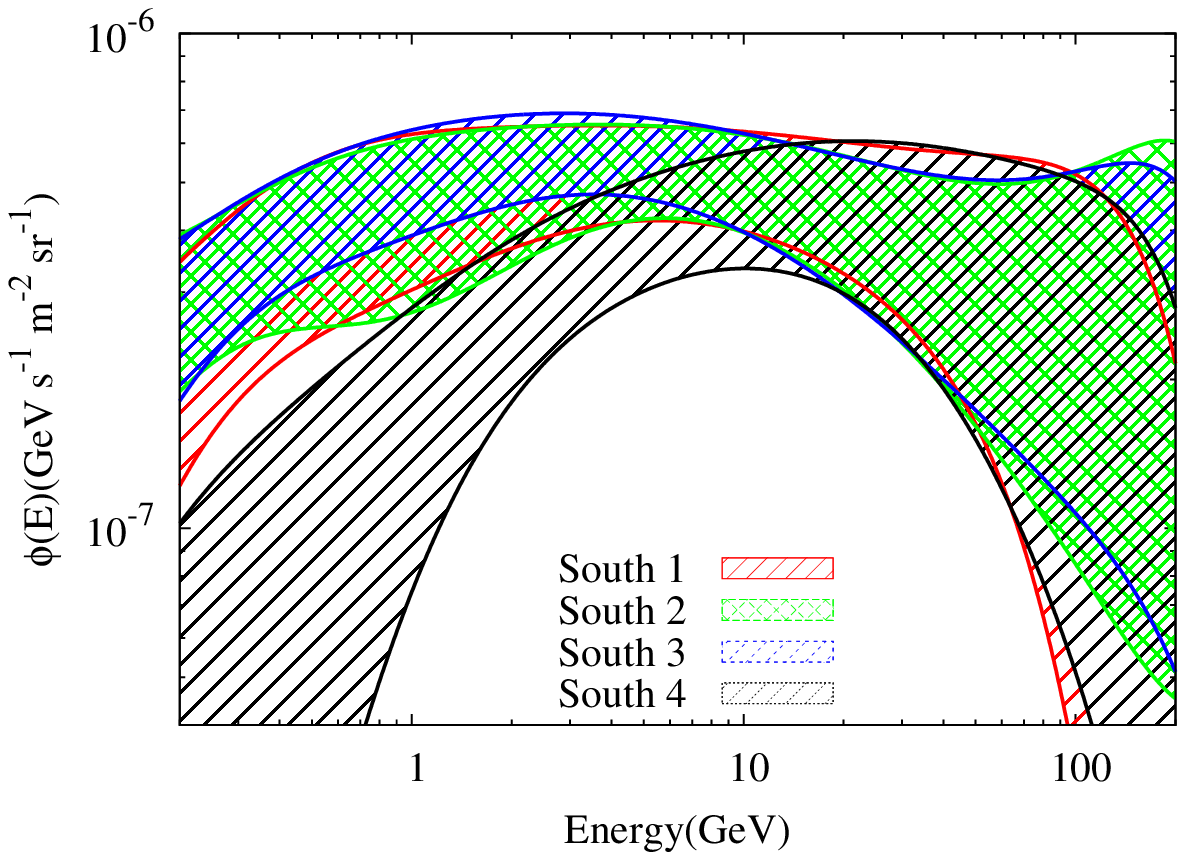}
\caption{The SEDs for the seven slices with all 64 templates we use in our analysis. The shaded areas span all  derived SEDs. The left panel is for the NFB while the right is for the SFB. }
\label{fig:sys}
\end{figure*}

\begin{figure*}
\centering
\subfigure[][{Lorimer distribution \citep{lorimer06}, $z_h=4, R_h=20, T_S=150$, and $E(B-V)_{cut} = 5$}]{\includegraphics[width=55mm,height=40mm,angle=0]{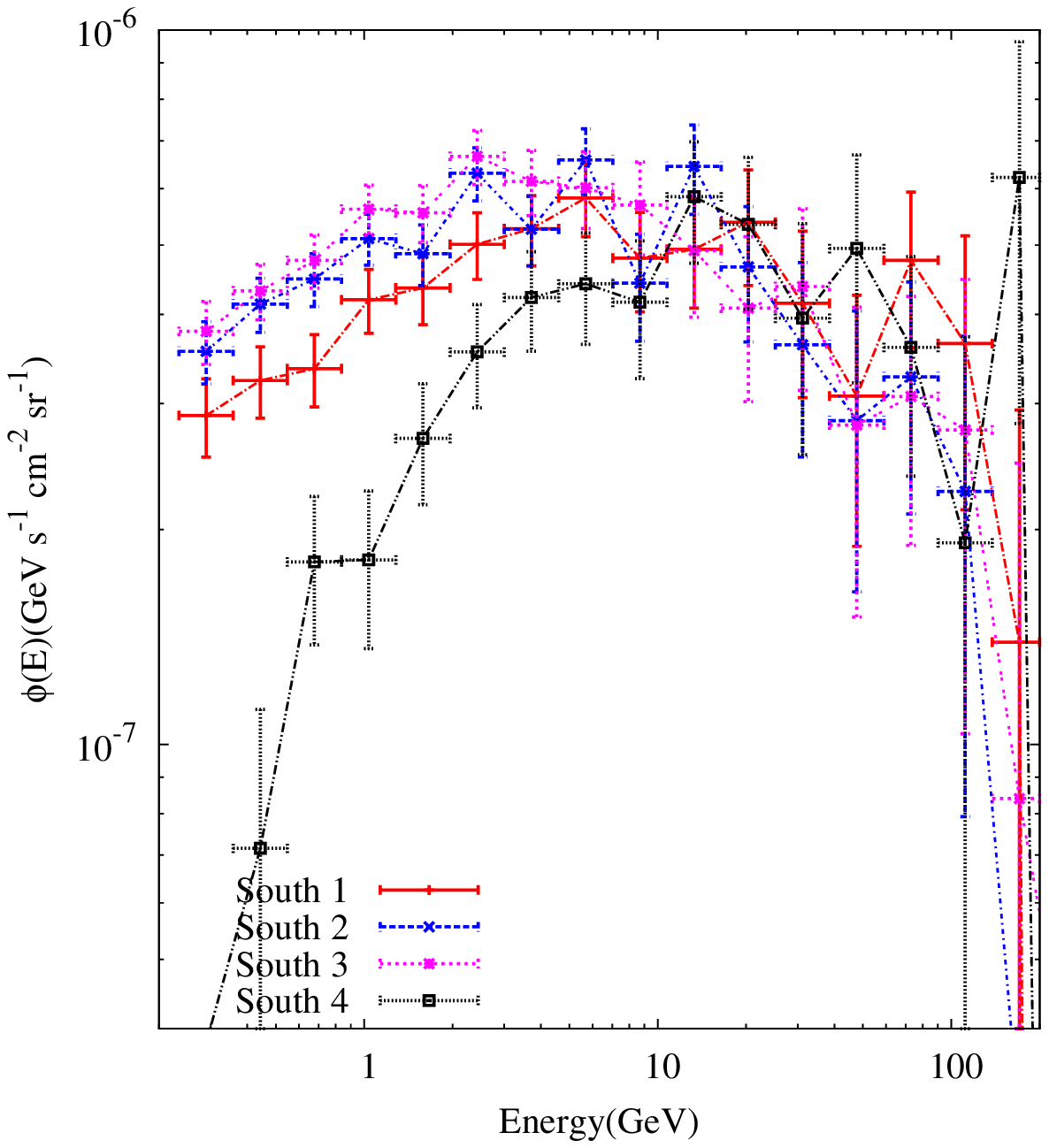}}~~~~~~~~~~
\subfigure[][{Lorimer distribution, $z_h=4, R_h=30, T_S=150$, and $E(B-V)_{cut} = 2$}]{\includegraphics[width=55mm,height=40mm,angle=0]{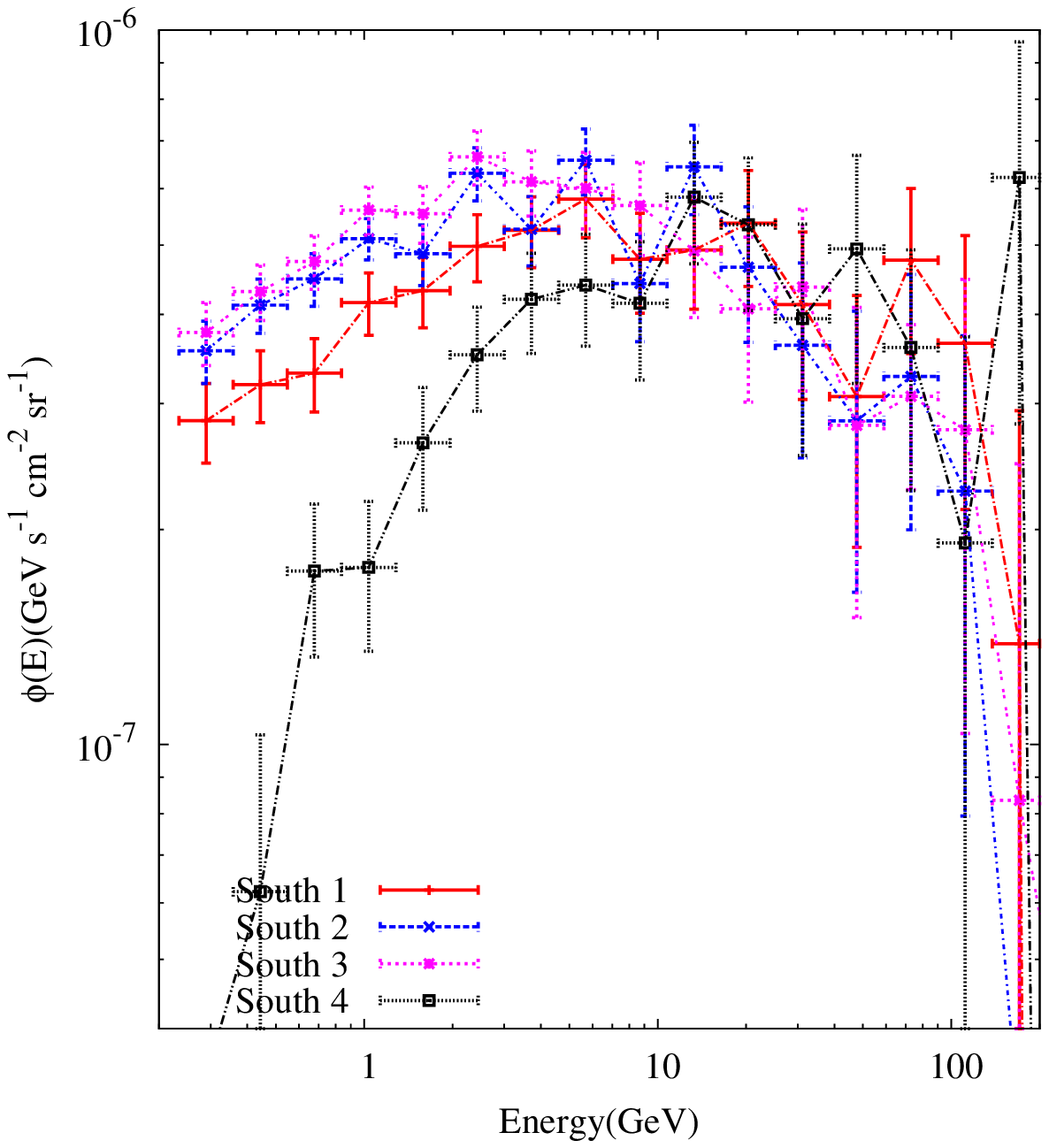}}\\
\subfigure[][{SNR distribution \citep{snrdis}, $z_h=6, R_h=20, T_S=150$, and $E(B-V)_{cut} = 2$}]{\includegraphics[width=55mm,height=40mm,angle=0]{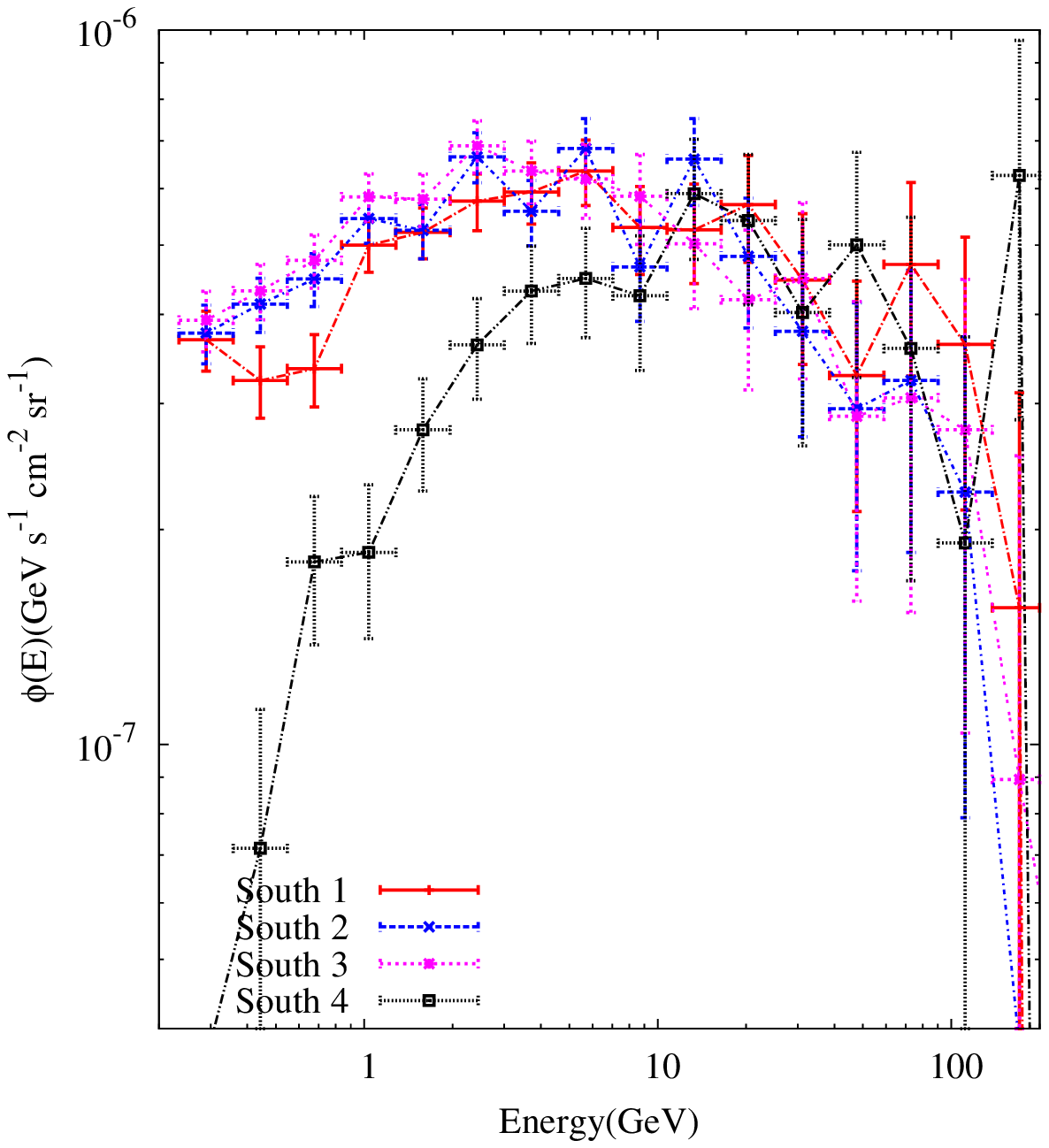}}~~~~~~~~~~
\subfigure[][{SNR distribution, $z_h=4, R_h=20, T_S=100000$, and $E(B-V)_{cut} = 2$}]{\includegraphics[width=55mm,height=40mm,angle=0]{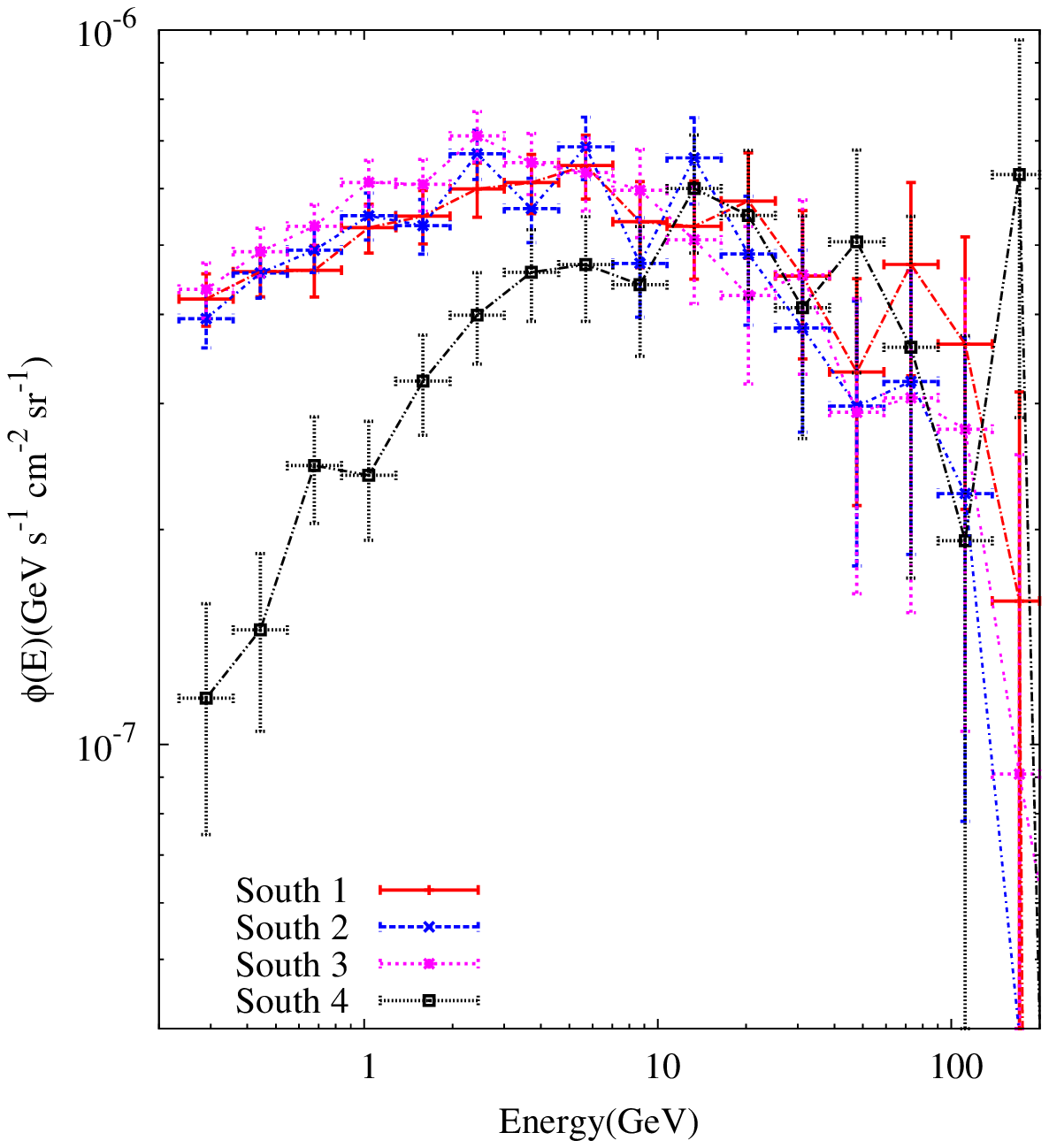}}\\
\subfigure[][{Yusifov distribution \citep{yusifov04}, $z_h=6, R_h=30, T_S=150$, and $E(B-V)_{cut} = 5$}]{\includegraphics[width=55mm,height=40mm,angle=0]{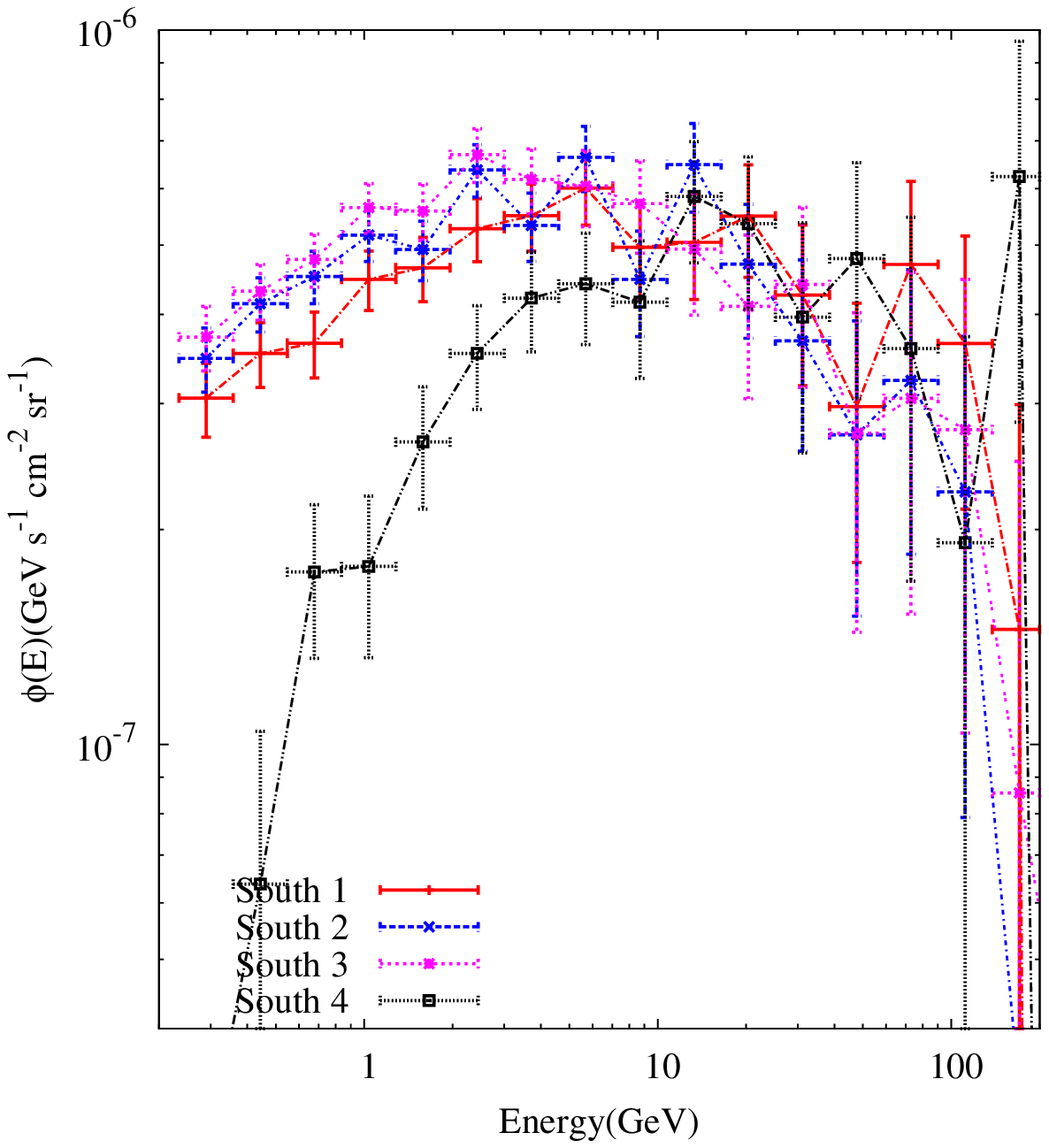}}~~~~~~~~~~
\subfigure[][{OBstars distribution \citep{obdis}, $z_h=4, R_h=20, T_S=150$, and $E(B-V)_{cut} = 2$}]{\includegraphics[width=55mm,height=40mm,angle=0]{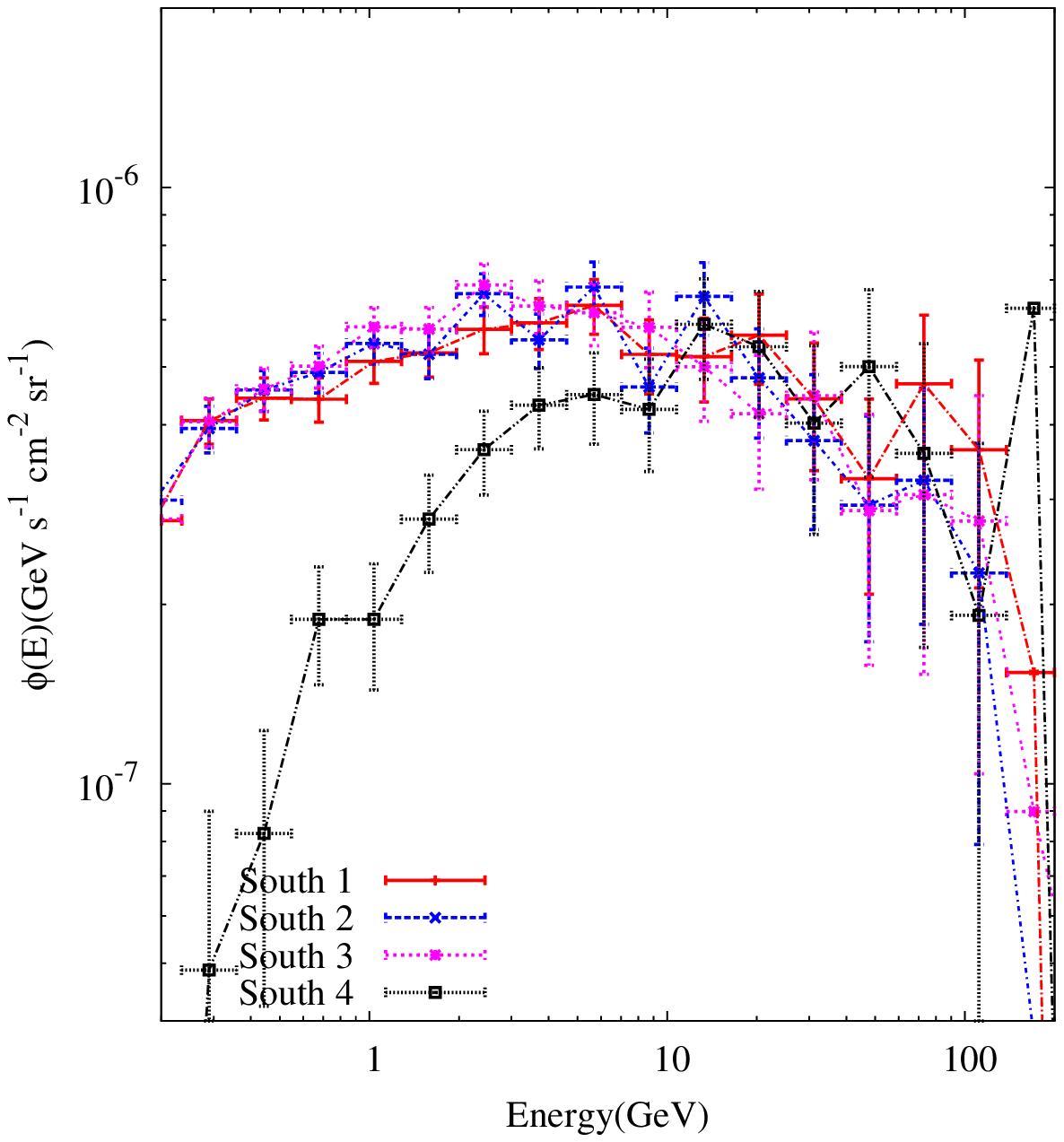}}
\caption{Six examples of SEDs for slices in the SFB. }
\label{fig:egs}
\end{figure*}

\begin{figure*}
\centering
\subfigure[][{Lorimer distribution, $z_h=4, R_h=20, T_S = 150$, and $E(B-V)_{cut} = 5$}]{\includegraphics[width=55mm,height=40mm,angle=0]{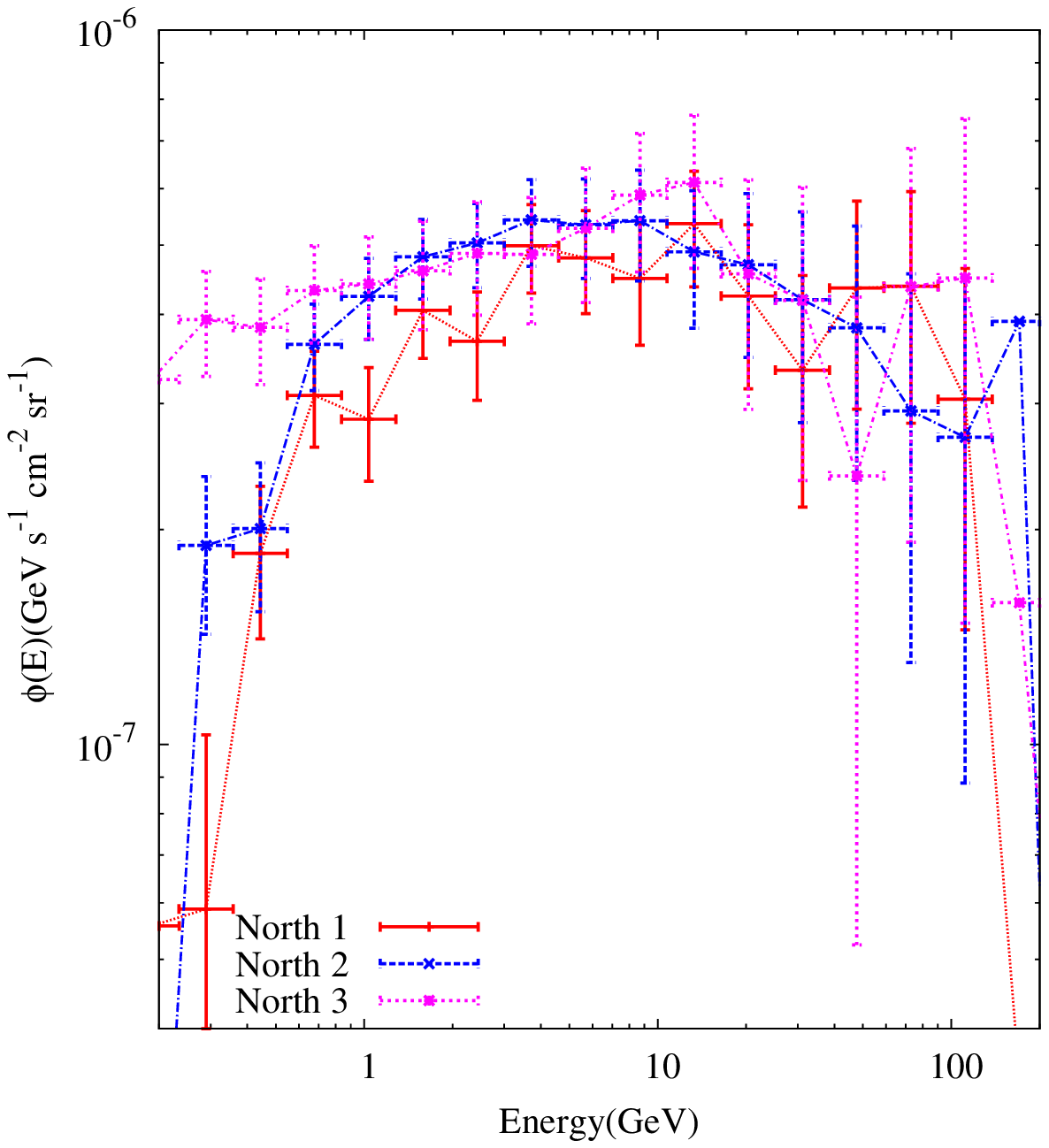}}~~~~~~~~~~
\subfigure[][{Lorimer distribution, $z_h=4, R_h=30, T_S=150$, and $E(B-V)_{cut} = 2$}]{\includegraphics[width=55mm,height=40mm,angle=0]{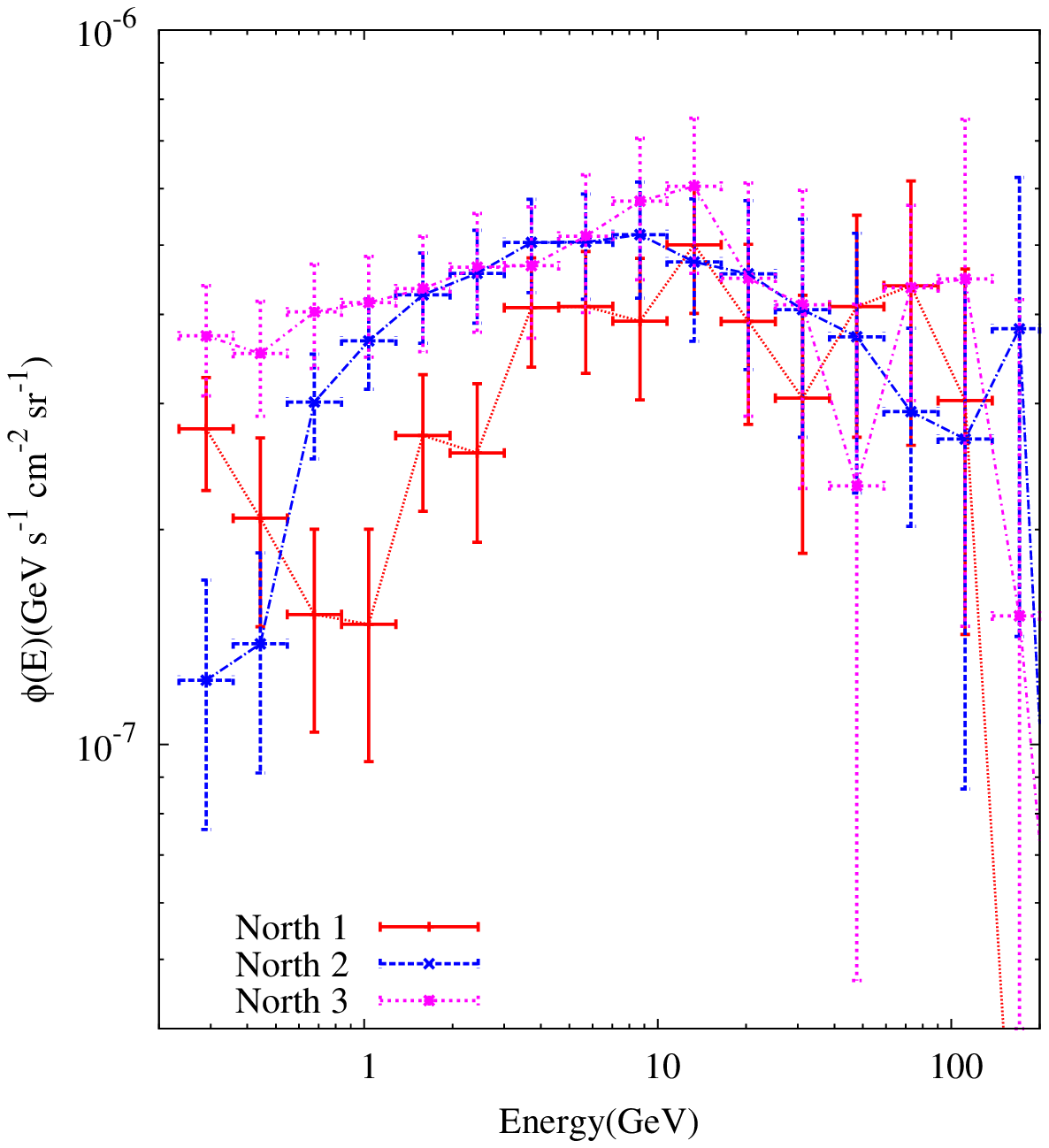}}\\
\subfigure[][{SNR distribution, $z_h=6, R_h=20, T_S=150$, and $E(B-V)_{cut} = 2$}]{\includegraphics[width=55mm,height=40mm,angle=0]{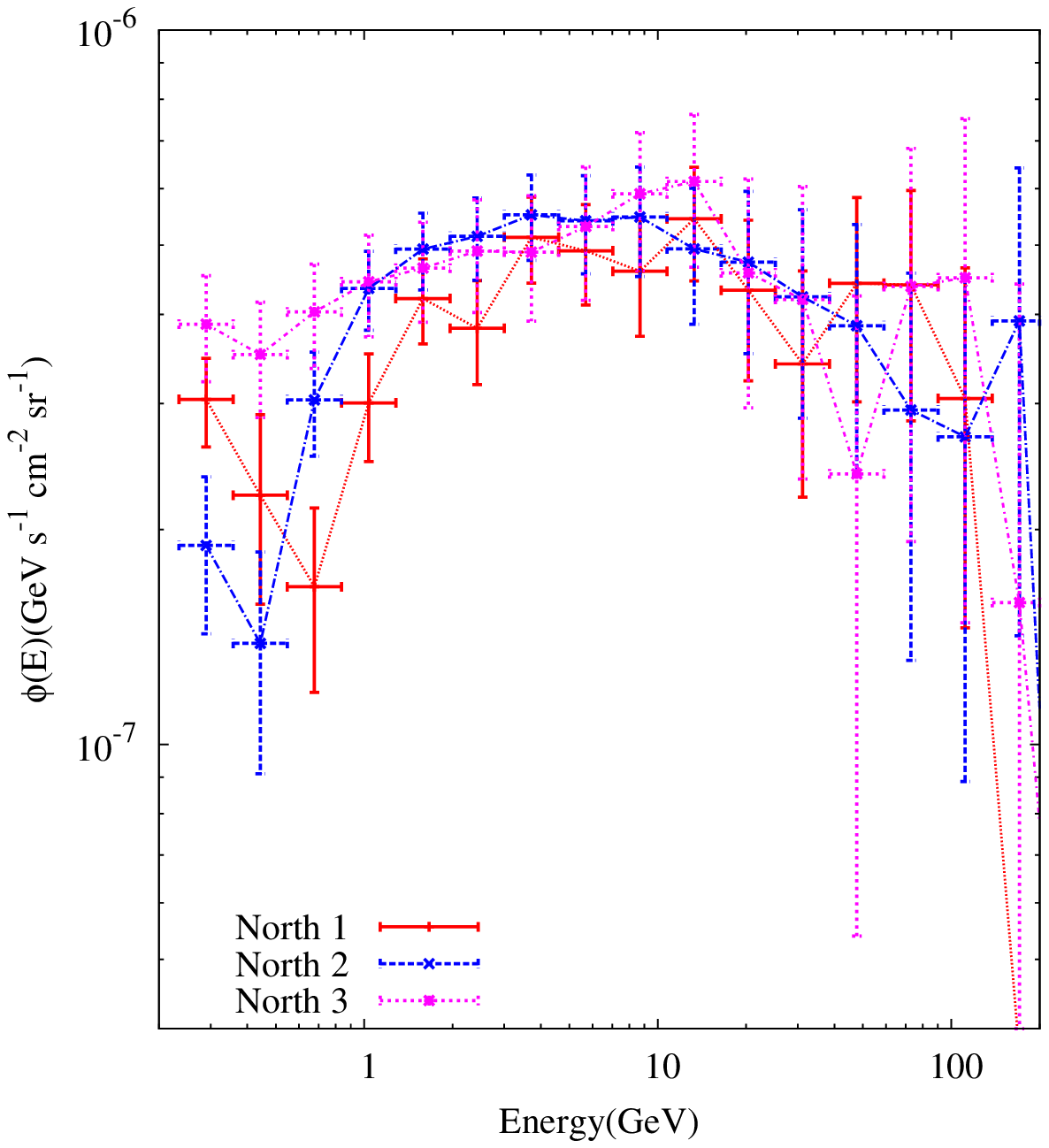}}~~~~~~~~~~
\subfigure[][{SNR distribution, $z_h=4, R_h=20, T_S=100000$, and $E(B-V)_{cut} = 2$}]{\includegraphics[width=55mm,height=40mm,angle=0]{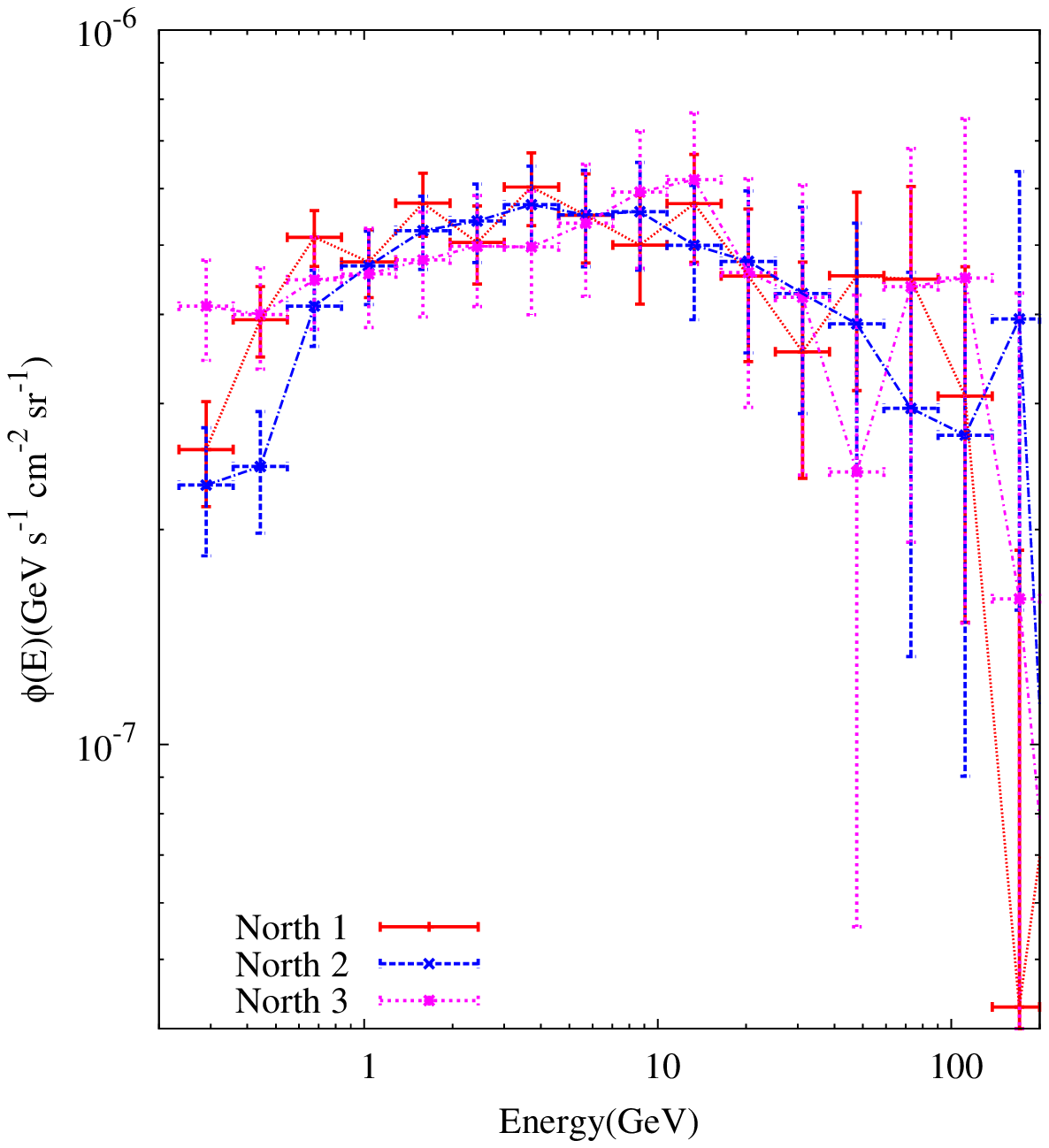}}\\
\subfigure[][{Yusifov distribution, $z_h=6, R_h=30, T_S=150$, and $E(B-V)_{cut} = 5$}]{\includegraphics[width=55mm,height=40mm,angle=0]{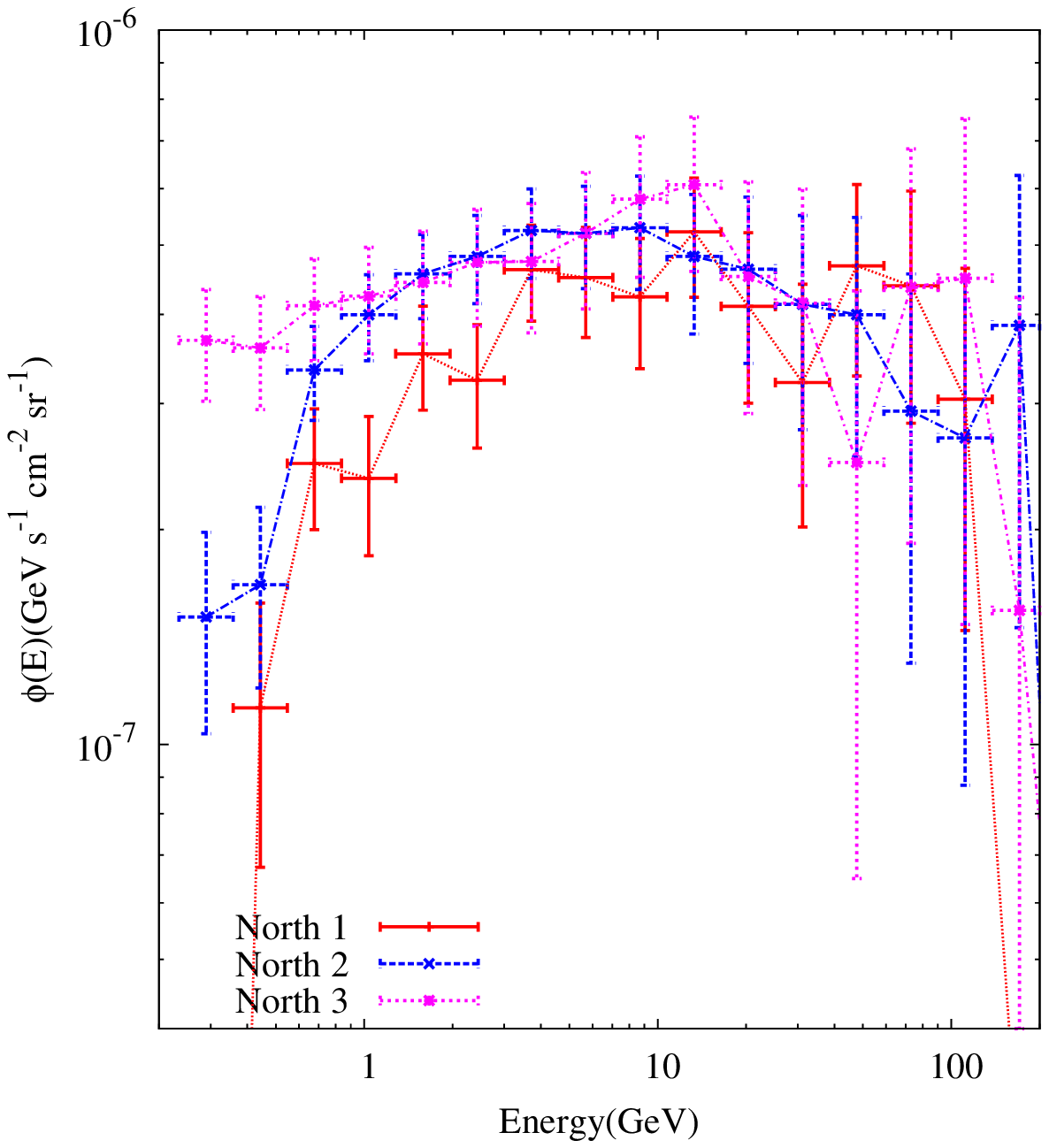}}~~~~~~~~~~
\subfigure[][{OBstars distribution, $z_h=4, R_h=20, T_S=150$, and $E(B-V)_{cut} = 2$}]{\includegraphics[width=55mm,height=40mm,angle=0]{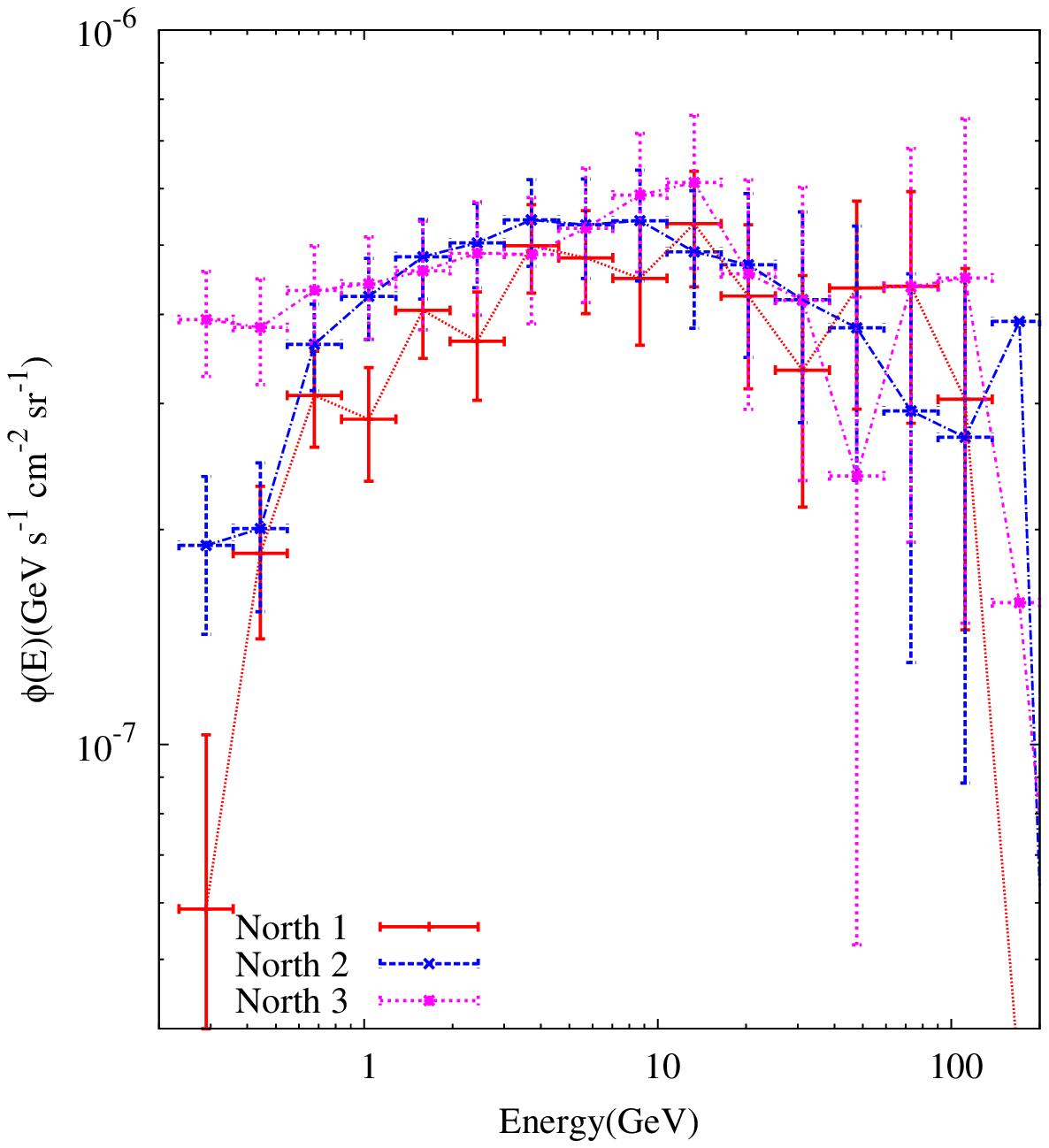}}
\caption{Six examples of SEDs for slices in the NFB. }
\label{fig:egn}
\end{figure*}

With the same procedure as described above, we find the SEDs in different slices with all chosen templates; the results are summarized in Fig.~\ref{fig:sys}. 
Because of the large systematic errors, we may not claim a deficit at low energies and high latitude in the NFB.  
However, in the SFB the low energy deficit of the top slice is  significant for every template.  
It should be noted that, due to the slight differences in overall flux normalisation for the different templates choices, 
differences of spectral shape are smeared out if we plot them all together as in Fig.~\ref{fig:sys} which, therefore, suggests a smaller difference between the top and bottom slices than exists in reality. 
In Fig.~\ref{fig:egs} we show 8 examples of SEDs for the slices in the SFB which give the extreme cases of the gamma ray flux.
From these individual examples the low energy deficit in the top slice is much more evident.  
Results for the northern slices are also shown in Fig.~\ref{fig:egn} which reveal the low energy part of the top slice has very large uncertainties. 
The large systematics in the NFB are likely due to the fact that it is partially coincident with the extended  Loop 1 structure
whose physical origin and exact morphology and spectral features are still uncertain.


\section{Discussion}

The gamma-ray residual maps obtained in the current study -- based on six years' {\it Fermi}-LAT data -- confirm that the surface brightness of  each bubble is  homogeneous at a gross level. 
This implies that the {\it volumetric} emissivity is not homogeneous, otherwise we would detect a higher brightness in the center of each bubble due to projection \citep{su10}. 
One possible explanation of this may be higher turbulence in the bubbles' edges generating more efficient particle acceleration or stronger confinement of high energy particles there with a resulting increase in the local gamma ray emissivity; alternatively, in a non-saturation hadronic model, higher gas densities towards the edges may have the same effect \citep{Crocker2013}.

Below we assume the low energy deficit  of the gamma ray spectrum at high latitudes found above is a real  effect and discuss its possible implications.  \\

To simplify our modeling, we assume each bubble is  a sphere with a radius 3.4 kpc.
The 2-d projection of each bubble can be approximated as a half sphere in the region $b>30^{\circ}$ for the NFB and $b<-30^{\circ}$ for the SFB. 
The center of each bubble is located at (0,0,$\pm$5kpc) in  Galactic coordinates. 
Assuming an {\it intrinsic} north-south symmetry, we only analyse the SFB and, to reveal most starkly the spectral change
with latitude, we consider only slice South 1 ($-33^\circ <b <-25^ \circ $) and South 4 ($-55^\circ <b <-47^ \circ $) in detail.  
Slice South 1 covers the center of the SFB.
We use the ISRF value in Galprop at  (0,0,5 kpc)  for the average value of the slice South 1 in the calculation, $w_{IR}= 0.18$ eV/cm$^3$ and $w_{opt}= $ 0.9 eV/cm$^3$. 
The height of South 4 is about 7 kpc, thus we adopt the ISRF energy density from Galprop at (0,0,7 kpc),  $0.7 $ eV/cm$^3$ for the optical component and $0.15 $ eV/cm$^3$ for the IR component. 
The IR  and optical photon fields are modeled as diluted blackbody (gray body) spectra with temperatures of 100 K and 5000 K, respectively.\\



\begin{figure*}
\centering
\includegraphics[width=70mm,angle=0]{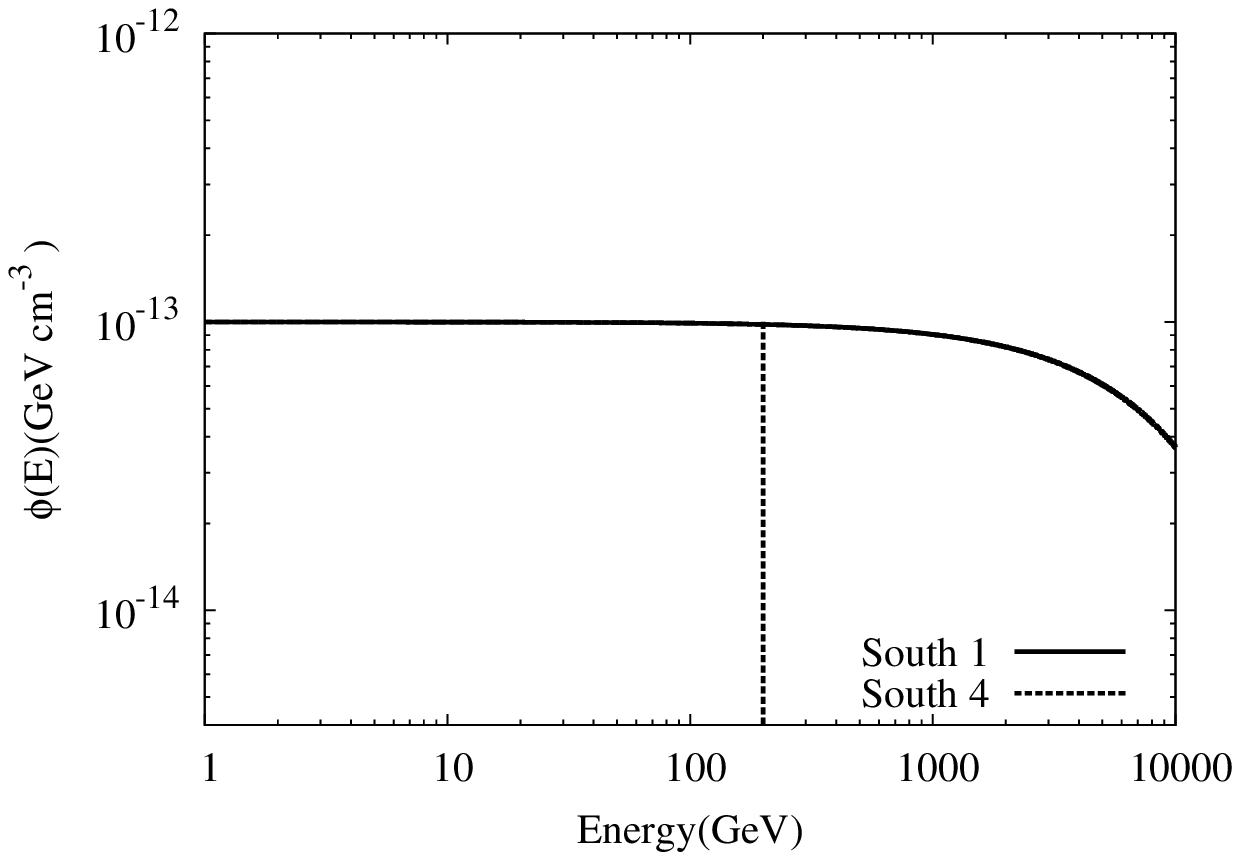}
\includegraphics[width=60mm,angle=0]{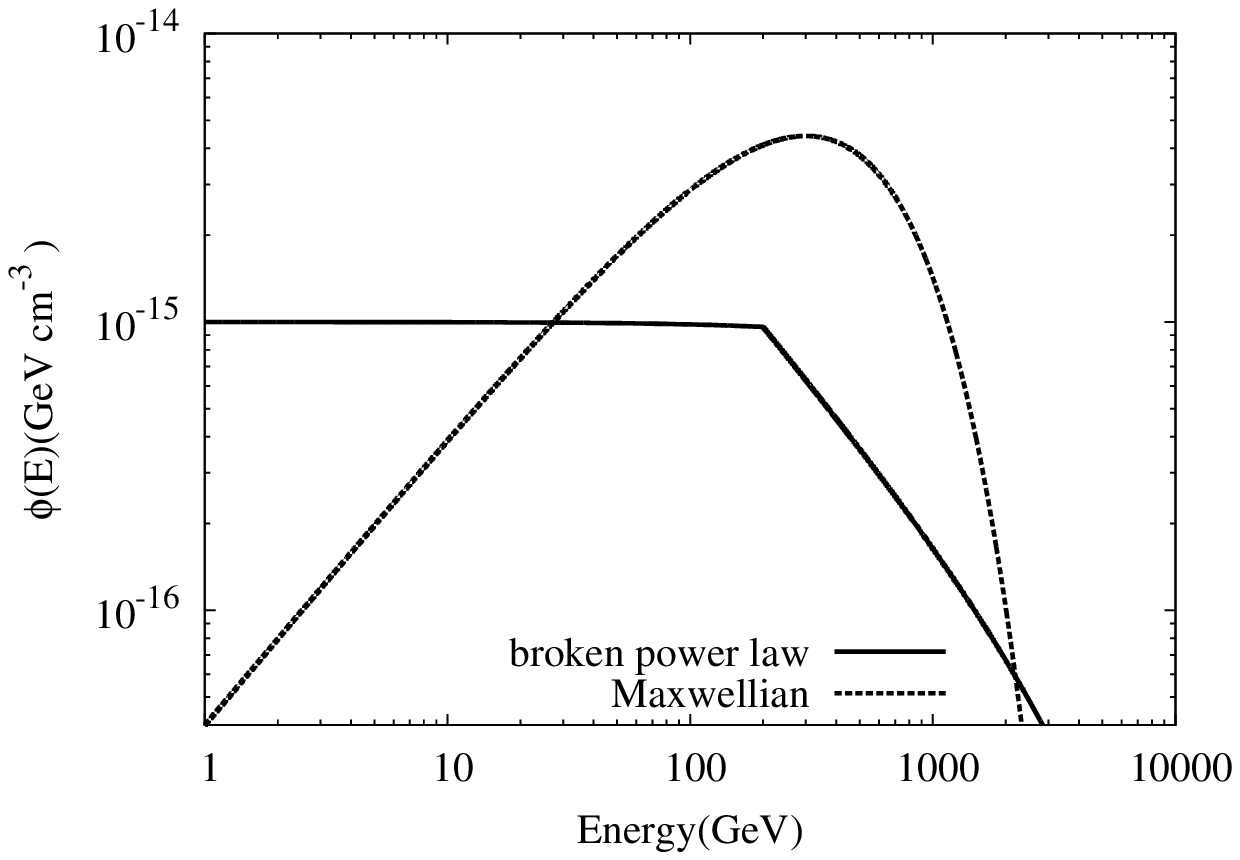}
\caption{Assumed CR proton (left) and electron (right) distributions in Slices 1 and 4 of the SFB}. 
\label{proton_spec}
\end{figure*}

\begin{figure*}
\centering
\includegraphics[width=50mm,angle=0]{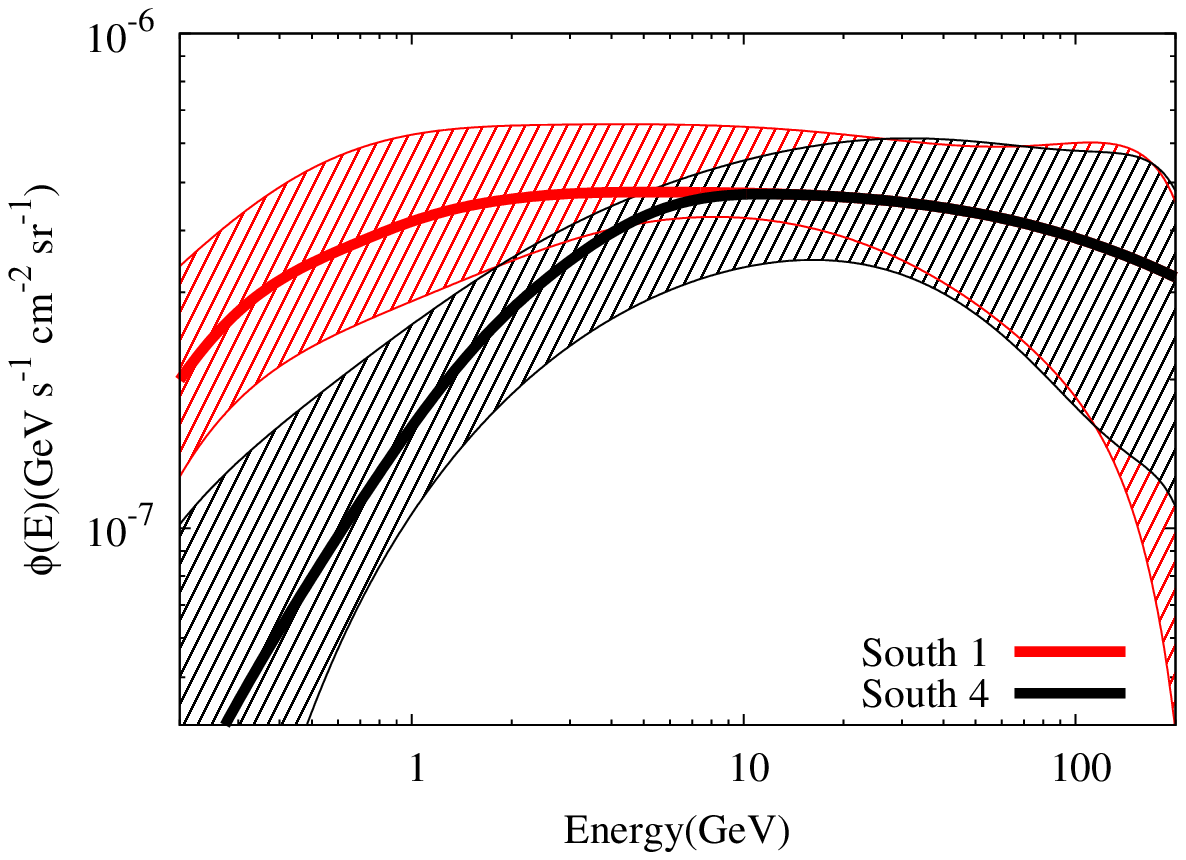}
\includegraphics[width=50mm,angle=0]{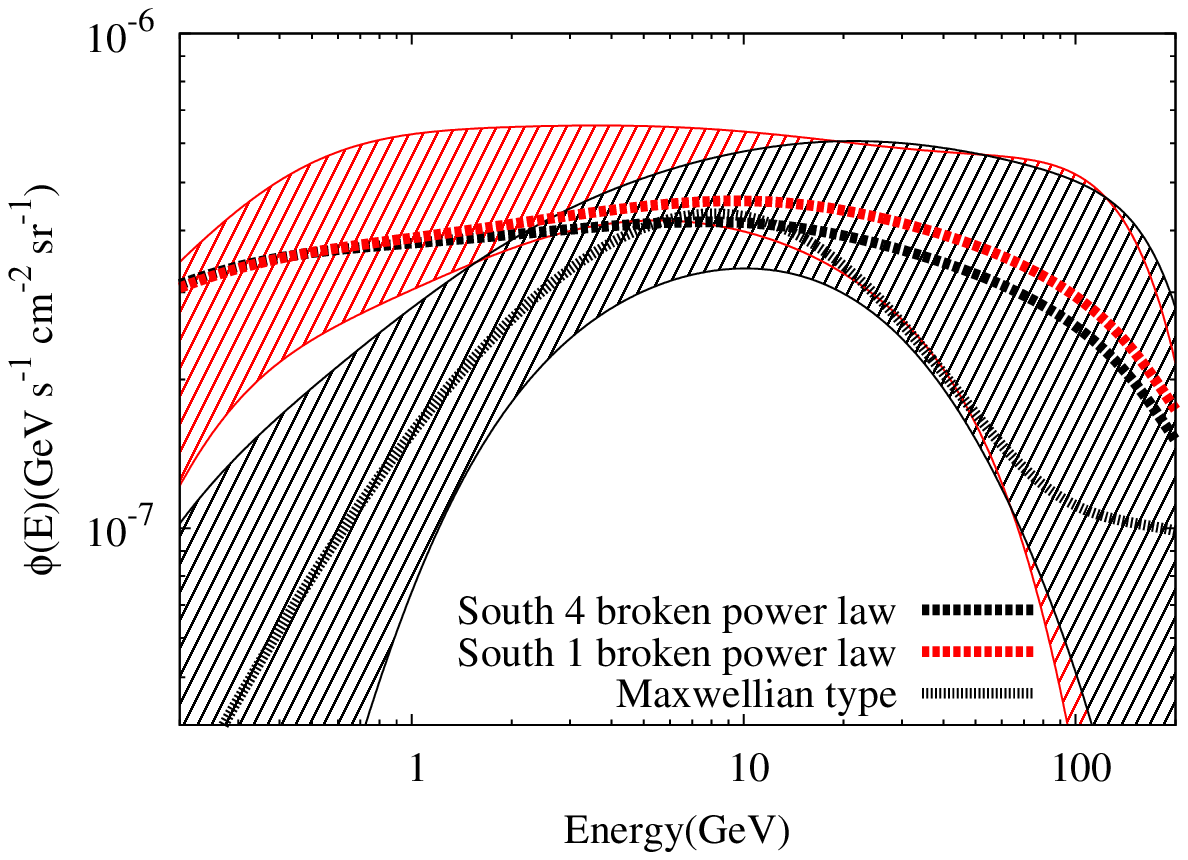}
\includegraphics[width=50mm,angle=0]{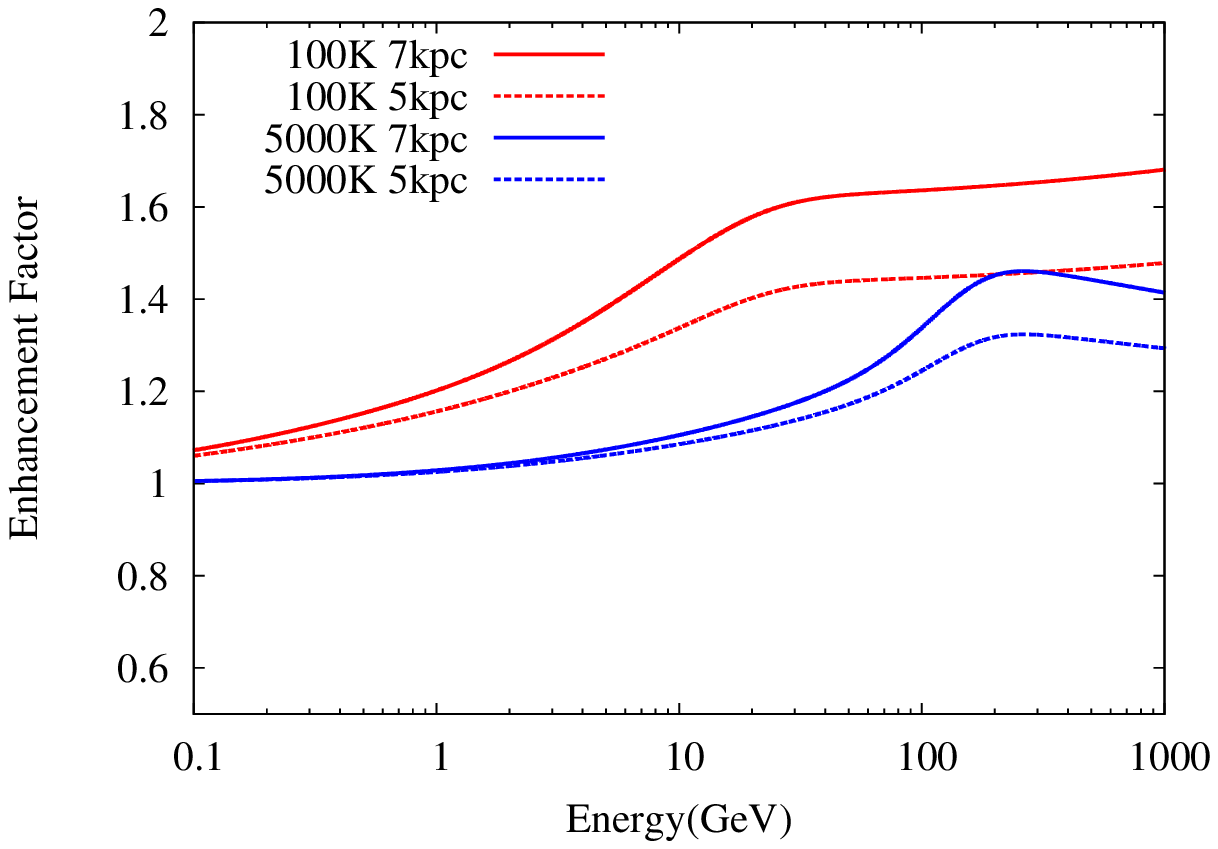}

\caption{ Left and Middle panel:  Fitting the FBs' SED with the hadronic model (left panel)  and the leptonic model (middle panel) described in the text. 
The red  curve shows the fitted spectrum and the red hatched region spans  all spectra resulting with from different templates for South 1;  blue curve and hatched region are for South 4.  
 For leptonic models,  for South 1 we assume the broken power law described in the text and the ISRF value given by GALPROP. 
 For South 4, the black curve is for the same broken power law with the ISRF floating while for the gray curve we use the ISRF in GALPROP at the position of South 4 and a Maxwellian type electron distribution with $E_{\rm bk}=300~\rm GeV$. 
 In our hadronic scenario  energy dependent diffusion effects are considered as  described in the text.  
 Right panel:  The enhancement factor due to anisotropic IC. The IR and Opt/UV components of the ISRF have effective  temperatures of $100$  K and $5000$ K, respectively.
  Results for $z=5~\rm kpc$ and $z=7~\rm kpc$ are shown (approximate heights of South 1 and South 4 above the Galactic plane).   }
\label{fig:pp}
\end{figure*}

In the hadronic scenario, motivated by the relative low energy deficit of the gamma ray flux in the top slice, we fit the SED in South 1 with a pure power law proton spectrum while for South 4 
we use a power law with a low energy break that we find needs to be at $20~\rm GeV$ to fit the data.
We assume the proton flux below the  break is zero for simplicity (see Fig.~\ref{proton_spec}). 
Our results are shown in Fig.~\ref{fig:pp}. 
The required low energy break may arise naturally due to   energy-dependent diffusion effects in a non-steady-state scenario: 
given South 4 is far from the (assumed) injection source in the plane, 
it may be that only high energy particles (which diffuse faster) have had time to travel there since a previous injection event or given the age of the structure. 
The formalism describing these energy-dependent diffusion effects can be found in \citet{aa96} where both  impulsive and continuous injection cases are described. 
The position of the low energy break can be estimated by equating the age of the bubble (or the time since a previous burst) with the diffusion time, i.e.,the beak energy is  $E_{\rm bk} \simeq (d^2/(D_{\rm 0} t ))^\frac{1}{\delta}$, 
where $d$ is the distance of the top slice from the injection source, $t$ is the age of the bubble (or burst) and $D(E)=D_{\rm 0}E^{\delta}$ is the energy dependent diffusion coefficient.  
If we assume  protons are injected at the Galactic center, $d$ is about $7~\rm kpc$.
Notice that in the kinetic equations which govern  proton diffusion only the combination $D_{\rm 0} \  t$ appears.
Thus any timescale can be obtained if we tune the diffusion coefficient $D_{\rm 0}$. 
If $D(E)$ takes a value similar to that in the Galactic plane, say  $4\times10^{28} $ cm$^2 $ s$^{-1}$ at 1 GeV and $\delta=0.3$, the age of the FBs should be $t \sim  10^8$ yr (cf.~\citealt{Crocker2013}). 
Alternatively, if there is continuous proton injection into the FBs over $10^{10}$~yrs we need a rather small diffusion coefficient, $D_0 \sim 10^{26}$ cm$^2$ s$^{-1}$ at 1 GeV, which is about 1/100 of that in the Galactic plane.

If we assume the injection is impulsive and  the ISM density is  $0.005$ cm$^{-3}$ the total energy in protons needed to light the Bubbles is of the order $10^{56} ~\rm erg$.
Note that if we assume the simple geometry mentioned above and a magnetic field $B \sim 10~\rm \mu G$ as suggested by \citet{carretti13}, then the energy densities of cosmic rays and magnetic fields are close to equipartition.  
A burst-like  injection event requires that the duration of the injection is much smaller than the age of the structure which is $t \sim  10^8$ yr (for $D_{\rm 0} = 4\times10^{28} $ cm$^2 $ s$^{-1}$ at 1 GeV), i.e., the duration of the injection event would have to be of order 
$10^7 ~\rm$ yr or less in this case.  
Then the injection rate is of the order $10^{41}$ erg/s, two orders of magnitude larger than the the X-ray luminosity of a X-ray reflection nebulae near the Galactic center \citep[e.g.,][]{ryu12}, but still only one thousandth of the Eddington Luminosity of Sgr $A^*$, $\sim 5\times 10^{44}~\rm erg/s$.

A possible concern for an energy-dependent diffusion scenario is that, in the high energy range for which the diffusion length is much larger than the distance to the source, the predicted proton density should be the same at all latitudes for the impulsive case, or scale as $1/R$ (where $R$ is the distance to the injection source) for the continuous case.
Thus, given the (expected) relatively smaller line of sight through the top of the SFB, one naively anticipates that the gamma ray flux  in South 4 is smaller than that in South 1. 
The observed data reveal a similar total flux in both slices at high energy, however. 
One possible resolution to this tension may be that the target gas density is higher at high latitudes.  
It is also likely that the geometry of the FBs is different from our simple  assumption of sphericity thereby entailing non-trivial projection effects.  \\
 
In a leptonic scenario the low energy deficit of gamma ray flux in the top slice also requires a corresponding low energy break in  the electron distribution (see Fig.~\ref{proton_spec} right).
This  can, again, be produced by several mechanisms. 
Firstly, the energy-dependent diffusion effects mentioned above also work for electrons in principle \citep[cf.][]{McQuinn2011}. 
However the much faster cooling of electrons relative to protons tends to prevent such effects playing an important role. 
To fit the high energy gamma ray data we need TeV electrons, whose cooling time scale is about $10^5~\rm yrs$ in the circumstances of the FBs. 
Thus, even if we assume very fast diffusion with the diffusion coefficient of $4\times 10^{29}$ cm$^2$/s at 1 GeV (and an index of 0.6), the diffusion distance of TeV electrons in $10^5~\rm$ yrs is only 2 kpc, significantly less than the distance of South 4 from the plane (which is more than 5 kpc). 
This implies high energy electrons accelerated at low latitudes can never reach South 4
and,
as a result, if we want to explain the radiation of the FBs in a leptonic scenario, the electrons should be accelerated {\it in situ}. \\

 The second problem related to electrons concerns  the interpretation of the very hard energy spectrum of radiation below $5~\rm GeV$.  Such a hard spectrum formally can be explained by assuming a low energy break in the electron 
spectrum.    However,  if the electrons are cooled,  no matter how hard the injection spectrum is, because of 
the radiative  ($dE/dt \sim E^2$ type) losses
the resulted {\it steady-state} distribution below the break in the injection spectrum,  will have have a standard,  
$E^{-2}$ power law-spectrum. 
The latter will result in IC gamma-ray spectrum  with a photon index 1.5 which  is still 
not sufficient to explain the observed gamma-ray spectrum  of South~4; below  $5~\rm GeV$
the photon index of the latter   is as small as $1$.  One can formally  overcome this problem assuming  
that the electrons stay uncooled in FBs.  However in order to reproduce the break in the gamma-ray spectrum, 
one would need a corresponding break at $100~\rm GeV$. 
The corresponding synchrotron cooling time at this energy,  $t \sim 8 \times 10^5 (\frac{100~ \rm GeV}{E})(\frac{B}{10~ \rm \mu G})^{-2} ~\rm yr$,  appears  significantly shorter than  the age of  FBs within any  reasonable  model  of the latter.  
Thus we  need very efficient continuous acceleration acceleration of electrons which would 
dominate over the rate of energy losses and in that way keep very hard the steady-state spectrum. 
This in principle can be realized  with stochastic acceleration which can produce Maxwellian  type {\it steady-state} energy 
distribution of electrons. Indeed adopting  an electron distribution like $N(E) \sim E^2 \ exp(-E/E_0)$ 
with  $E_0= 300~\rm$ GeV,  we can have a reasonable fit to the detected gamma-ray spectrum  in South 4,
as it is shown in Fig.~\ref{fig:pp}.  For comparison, we also  show  IC emission at the position of 
South 1 and South 4 by adopting a  broken power law electron distribution with an index $2$ below the break at $200~\rm GeV$ and $3$ above.  The results fit the  SED of South 1 well, but 
fail to fit that of South 4, which are also shown in Fig.~\ref{fig:pp}.

Anisotropy effects may, in principle, increase the IC  gamma ray flux at high latitudes since most  of  seed photons come from regions  in  the Galactic plane;   thus gamma-rays we detect at high latitudes
are  a result of (almost)  head-on collisions. 
To probe  this effect we use the formalism described in \citet{aniic}. 
The enhancement factor is defined by the flux ratio of the anisotropic IC  to the isotropic IC  assuming the 
the same ISRF energy density. 
Results for the positions of South 1 and South 4 are shown in Fig.~\ref{fig:pp}. 
We can see that although the anisotropic IC effect may  cause a spectral a hardening, 
however  the difference between South~1 and South~4 is rather  small, less than $10$\%. It appears not 
sufficient to explain the  very hard spectrum of South~4. 

\section{Summary}
In this paper we re-analyze the {\it Fermi}-LAT data covering the FB region. 
With an improved  instrumental response function and more data, we are able to extend the previously obtained spectrum to lower energies. 
Furthermore we divide  each bubble into  slices to investigate  possible variation of the spectrum with latitude. 
Given the improved data, for the first time, we can determined robustly that the spectrum of the top of South Bubble drops appreciably at low energies relative
to the spectrum determined at lower latitudes.
We also show that the morphology of the South Bubble is  energy dependent; at high energy the  structure is relatively more extended to both Galactic south and west.

We have also investigated the influence of different choices of background model on the  results of our analysis. 
We found that background models may significantly alter the apparent gamma-ray signal from the FBs.
Nevertheless, the spectral hardening with latitude in the South Bubble remains a robust result.
In neither Bubble do we find a spectral {\it steepening} with latitude.

The relative suppression of the low energy gamma ray spectrum may be explained within a hadronic model for the FBs wherein
energy-dependent diffusion leads to a relative deficit of low energy protons at high latitudes.
Specifically, if  protons are injected in the plane, the finite age of the FBs may only allow  high energy protons to propagate to high latitude thereby predicting a gradual hardening
of the  gamma ray spectrum with latitude. 

In leptonic models the fast cooling of electrons means they cannot move too far from their accelerators and distributed acceleration 
inside the FBs seems to be favored. 
For the  case of stochastic acceleration the maximal electron energy might be expected to be $1~\rm TeV$, implying the optical/UV and IR component of the ISRF may play an important role in  gamma ray production inside the FBs. 
However, the attenuation of these components of the ISRF at high latitudes is, at least naively, in conflict with the observed hardening of the gamma ray spectrum in the highest slice of the South Bubble unless a very specific electron spectrum evolution is realized.  \\

\bibliography{slicev14}
\bibliographystyle{aa}

\end{document}